\pdfoutput=1

\documentclass[acmsmall,screen=true,review=false]{acmart}

\usepackage{multirow}
\usepackage{multicol}
\usepackage{bbm}
\usepackage[inline]{enumitem} 
\usepackage{amsmath}
\usepackage{mathtools}
\usepackage{hyperref}       
\usepackage{url}            
\usepackage{csquotes}
\usepackage{natbib}
\usepackage{graphicx}
\usepackage{wrapfig}
\usepackage{tabularx}
\usepackage[skip=2pt]{caption}
\usepackage{subcaption}

\usepackage{numprint}
\AtBeginDocument{%
  \providecommand\BibTeX{{%
    \normalfont B\kern-0.5em{\scshape i\kern-0.25em b}\kern-0.8em\TeX}}}

\newcommand{\heading}[1]{\vspace*{1mm}\noindent\textbf{#1.}}

\setcopyright{rightsretained}
\copyrightyear{2024}
\acmYear{2024}
\acmDOI{}
\acmJournal{TOIS}
\acmVolume{}
\acmNumber{}
\acmArticle{}
\acmMonth{2}

\ccsdesc[500]{Information systems~Retrieval models and ranking}

\keywords{Generative retrieval, Continual learning, Knowledge-intensive language tasks}

\received{}

\author{Jiafeng Guo} 
\orcid{0000-0002-9509-8674}
\email{guojiafeng@ict.ac.cn
} 

\author{Changjiang Zhou}
\orcid{0009-0000-0005-9465}
    \email{zhouchangjiang23s@ict.ac.cn
}  

\author{Ruqing Zhang}
\orcid{0000-0003-4294-2541}
    \authornote{Ruqing Zhang is the corresponding author. Research conducted when the author was at the University of Amsterdam.}
    \email{zhangruqing@ict.ac.cn
}

\author {Jiangui Chen}
\orcid{0000-0002-6235-6526}
\email{chenjiangui18z@ict.ac.cn
}    

\affiliation{
  \institution{
  Institute of Computing Technology, Chinese Academy of Sciences; 
  University of Chinese Academy of Sciences}
  \streetaddress{NO. 6 Kexueyuan South Road, Haidian District}
  \city{Beijing}
  \country{China}
  \postcode{100190}
}

\author{Maarten de Rijke}
\orcid{0000-0002-1086-0202}
\email{m.derijke@uva.nl}

\affiliation{
  \institution{University of Amsterdam}
  \city{Amsterdam}
  \country{The Netherlands}
}

\author{Yixing Fan} 
\orcid{0000-0003-4317-2702}
\email{fanyixing@ict.ac.cn
} 

\author{Xueqi Cheng}
\orcid{0000-0002-5201-8195}
\email{cxq@ict.ac.cn
}

\affiliation{
  \institution{
  Institute of Computing Technology, Chinese Academy of Sciences; 
  University of Chinese Academy of Sciences}
  \streetaddress{NO. 6 Kexueyuan South Road, Haidian District}
  \city{Beijing}
  \country{China}
  \postcode{100190}
}

\renewcommand{\shortauthors}{Guo et al.}

\begin{document}

\title{CorpusBrain++: A Continual Generative Pre-Training Framework for Knowledge-Intensive Language Tasks}

\begin{abstract}
Knowledge-intensive language tasks (KILTs) typically require retrieving relevant documents from trustworthy corpora, e.g., Wikipedia, to produce specific answers.
Very recently, a pre-trained generative retrieval model for KILTs, named CorpusBrain, was proposed and reached new state-of-the-art retrieval performance.
However, most existing research on KILTs, including CorpusBrain, has predominantly focused on a static document collection, overlooking the dynamic nature of real-world scenarios, where new documents are continuously being incorporated into the source corpus. 
To address this gap, it is crucial to explore the capability of retrieval models to effectively handle the dynamic retrieval scenario inherent in KILTs.

In this work, we first introduce the continual document learning (CDL) task for KILTs and build a novel benchmark dataset named KILT++ based on the original KILT dataset for evaluation. 
Then, we conduct a comprehensive study over the use of pre-trained CorpusBrain on KILT++.  
Unlike the promising results in the stationary scenario, CorpusBrain is prone to catastrophic forgetting in the dynamic scenario, hence hampering the retrieval performance.
To alleviate this issue, we propose CorpusBrain++, a continual generative pre-training framework that enhances the original model in two key aspects:
\begin{enumerate*}[label=(\roman*)]
    \item We employ a backbone-adapter architecture: the dynamic adapter is learned for each downstream KILT task via task-specific pre-training objectives; the backbone parameters which are task-shared are kept unchanged to offer foundational retrieval capacity.  
    \item We leverage the experience replay strategy based on exemplar documents that are similar to new documents, to prevent catastrophic forgetting of old documents.
\end{enumerate*}
Empirical results demonstrate the significant effectiveness and remarkable efficiency of CorpusBrain++ in comparison to both traditional and generative IR methods.
\end{abstract}

\maketitle
\section{Introduction}

Knowledge-intensive language tasks (KILTs) refer to a series of language-related tasks that require access to  external knowledge sources such as Wikipedia for accurate answer generation~\cite{petroni2020kilt}. 
In current mainstream approaches, a two-step process is commonly employed \cite{chen2017reading,kwiatkowski2019natural,yang2018hotpotqa}, consisting of a retriever and a reader.  
The retriever aims to retrieve relevant documents from large, external knowledge sources, while the reader is meant to synthesize the retrieved information to generate accurate and correct answers to the initial query. 
Thanks to the emergence of large-scale pre-trained generative language models \cite{lewis2020bart,raffel2020exploring}, the reader component has seen remarkable advances recently.  
The retriever component has primarily leaned on conventional discriminative methods \cite{guo2020deep}, failing to fully capitalize on the potential advantages offered by generative models.

Generative retrieval (GR) has recently been proposed as an alternative retrieval paradigm~\cite{metzler2021rethinking}. 
In GR, the retrieval process is formalized as a sequence-to-sequence (Seq2Seq) learning problem, i.e., directly establishing a mapping from a query to its relevant document identifiers (docids). 
In essence, a single generative model is utilized to encode all information about the corpus into model parameters, allowing for end-to-end optimization and facilitating the alleviation of computational costs. 
As a result, GR stands out as a highly promising paradigm for retrieval in KILTs when compared to traditional discriminative methods. 
Specifically, previous research has investigated direct applications of pre-trained generative language models in the natural language processing (NLP) field, such as BART~\cite{lewis2020bart} and T5~\cite{raffel2020exploring}, to the KILT retrieval task~\cite{tay2022transformer,bevilacqua2022autoregressive,chen2022gere}. 
This approach involves initializing the model parameters with pre-trained generative models and subsequently fine-tuning them using golden query-docid pairs in downstream KILTs, which has demonstrated notable performance improvements in retrieval tasks.

Beyond the direct application of existing pre-trained generative models designed for NLP, there have been some pioneer studies on constructing generative pre-training tasks tailored for the KILT retrieval task.
The underlying hypothesis is that leveraging pre-training tasks that more closely resemble the relevance  relationship between queries and documents in downstream KILT tasks can yield better retrieval performance~\cite{glass2020span, ke2020sentilare,zhang2020pegasus}.
A latest and representative contribution following this research domain pertains to CorpusBrain~\cite{chen2022corpusbrain}, whose results reported on the KILT leaderboard\footnote{\url{https://eval.ai/challenge/689/leaderboard}} showcase new state-of-the-art performance, surpassing strong baselines. 
The key idea of CorpusBrain is to construct pre-training data consisting of positive pairs of queries and docids that encompass various semantic granularities in downstream tasks. 
Subsequently, a transformer-based~\cite{vaswani2017attention} encoder-decoder architecture is pre-trained by maximizing the likelihood of the output sequence with a standard Seq2Seq objective.

\begin{figure}[t]
  \centering 
  \vspace{-3mm} 
  \includegraphics[width=\linewidth]{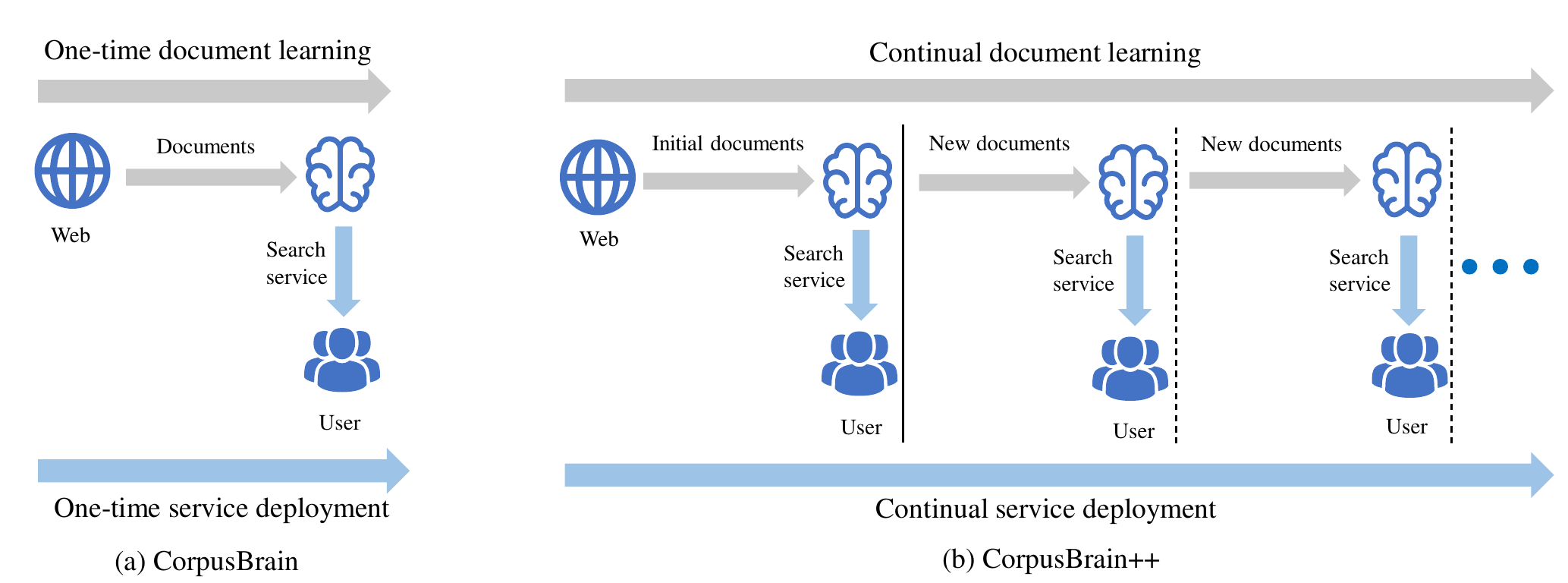}
  \caption{Comparison of CorpusBrain and CorpusBrain++. CorpusBrain can solely support one-time document learning and service deployment, without the ability to assimilate new documents and  dynamically update the knowledge base. Beyond CorpusBrain, the dynamic CorpusBrain++ can support continual document learning and service deployment to adapt to evolving corpus in the realistic scenario.}
  \label{illustration}
    \vspace{-3mm}
\end{figure}

The majority of prior retrieval models developed for KILTs, including CorpusBrain, have primarily focused on the scenario of stationary knowledge sources, as shown in Figure \ref{illustration} (a): whenever they finish learning, they would remain unchanged when used in practice. 
In contrast to the static assumption, the accrual of knowledge over time is a ubiquitous phenomenon in most real-world scenarios, giving rise to new documents added to the underlying knowledge source.
For instance, the open-source Wikipedia has experienced exponential growth in the number of documents\footnote{\url{https://en.wikipedia.org/wiki/Wikipedia:Size_of_Wikipedia}} since its inception in 2001~\cite{almeida2007evolution}, and new entities emerge following following, in many cases, the news cycle~\citep{graus-birth-2018}. 
Therefore, to ensure that a generalist chatbot remains up-to-date and well-informed in the face of this ever-changing information landscape, it is imperative for the chatbot to consistently expand its knowledge coverage. 
In traditional dense retrieval methods \cite{karpukhin2020dense,zhao2022dense}, the process of incorporating new documents into the retrieval system is relatively straightforward; the encoded representations of the incremental documents can be directly added to an explicit external index, without requiring updates to the retrieval model itself. 
However, in the case of the state-of-the-art CorpusBrain model, the dynamic retrieval scenario poses a more significant challenge, mainly due to the use of an implicit parameterized index.
Hence, it is of critical importance to investigate the ability of CorpusBrain to continuously accommodate the inclusion of new documents.

In this work, as illustrated in Figure \ref{illustration} (b), we make the first attempt to concentrate on the dynamic retrieval scenario for KILTs. 
Firstly, we formally define the continual document learning (CDL) task for KILTs and outline the corresponding evaluation metrics. 
To facilitate fair and quantitative comparisons between different models in their ability to tackle the CDL task, we then introduce a novel benchmark dataset named KILT++, which is constructed by splitting the original KILT dataset~\cite{petroni2020kilt} into distinct sessions to simulate the continual addition of new documents.
Subsequently, we assess the performance of two unsophisticated variants of the off-the-shelf pre-trained CorpusBrain model on the newly constructed KILT++ dataset, i.e., the direct insertion approach and the sequential pre-training approach. 
Our empirical findings confirm that, unlike the promising results achieved in the static scenario, CorpusBrain is vulnerable to catastrophic forgetting and inadequate in the ability to effectively and efficiently address the dynamic retrieval scenario.

To tackle the non-trivial CDL task, 
we propose a continual generative pre-training framework for KILTs, namely, CorpusBrain++ (CorpusBrain + new documents), to adapt CorpusBrain to the dynamic nature of constantly evolving corpora in practice. 
CorpusBrain++ targets to enable accurate retrieval of both old and new documents for queries, without catastrophic forgetting of previous knowledge. 
To achieve this objective, we need to address two main challenges:
\begin{enumerate}[label=(\roman*)]
    \item How to expeditiously learn specific retrieval capacity for each KILT task as the corpus constantly evolves? 
    \item How to prevent catastrophically forgetting the retrieval capacity already learned?
\end{enumerate}

\noindent%
Specifically, we have advanced beyond the original CorpusBrain model in two key directions to solve the aforementioned challenges: 
\begin{enumerate}[label=(\roman*)]
   \item In CorpusBrain++, we leverage a backbone-adapter architecture, wherein a dedicated adapter is employed for each downstream task to allow for capturing task-specific characteristics. 
   The fixed backbone component serves as long-term memory to retain fundamental retrieval capacity, while the dynamic adapter component serves as short-term memory to rapidly learn incremental documents.
   To enable continual pre-training of the task-specific adapters,
   we further design a pre-training task specifically tailored for each individual task.
   \item To avoid catastrophic forgetting of old documents, we use experience replay based on exemplar documents. We revisit old documents that are semantically similar to the incremental documents and apply the specific pre-training tasks for both the newly-arrived documents and the revisited ones.
\end{enumerate}

\noindent%
We assess the performance of CorpusBrain++ on the constructed KILT++ dataset.
The empirical results demonstrate that CorpusBrain++ excels in efficiently and effectively handling the CDL task for KILTs. 
Further ablation studies are conducted, revealing the effectiveness of each individual component within the CorpusBrain++ architecture.
Moreover, through our experimental analysis, we confirm that CorpusBrain++ successfully mitigates the occurrence of catastrophic forgetting of previously encountered documents and showcases the capability of positive forward knowledge transfer. 
Finally, we also analyze the effectiveness-efficiency trade-off of our method and conduct a case study to further reveal the inner workings of our method.

\section{Continual document learning for KILTs} \label{setting}

Here, we first introduce the continual document learning (CDL) task for KILTs and then describe the constructed KILT++ benchmark dataset, and finally elucidate the corresponding evaluation metrics to assess the CDL task for KILTs.

\subsection{Task formulation}

Assume we have a large-scale base document set (i.e., Wikipedia articles) $\mathcal{D}_0$ and sufficiently many labeled query-document pairs $\mathcal{R}_0$ in downstream KILT tasks. 
Here, $\mathcal{R}_0$ contains all the labeled datasets in different KILT tasks, specifically including fact checking, entity linking, slot filling, open-domain question answering (QA), and dialogue.

In the CDL task for KILTs, we assume there exist $T$ batches of new documents $\{\mathcal{D}_1$, \dots, $\mathcal{D}_t$, \dots, $\mathcal{D}_T\}$, which arrive in a sequential manner as the time session grows. 
In any session $t\ge1$, the corresponding labeled KILT data $\mathcal{R}_t$ is not available, i.e., $\mathcal{D}_t$ is only composed of newly encountered documents $\{d_t^1, d_t^2, \dots\}$ without labeled queries relevant to these documents. 
Let the retrieval model after the $t$-th update be $\mathcal{M}_t$ and the model parameters be $\Theta_t$.
For session $t$, the training objective of CDL for KILT can be defined as updating $\Theta_{t-1}$ to $\Theta_{t}$ via the new document set $\mathcal{D}_t$ and previous datasets  $\{\mathcal{D}_0,\dots,\mathcal{D}_{t-1}\}$, such that $\mathcal{M}_t$ can simultaneously retrieve relevant documents from previously and
newly arrived documents $\{\mathcal{D}_0,\dots,\mathcal{D}_t\}$. 
To assess the retrieval performance of $\mathcal{M}_t$, we employ the test set $\mathcal{Q}_i^{\delta}, i\le t$, where $\delta$ denotes the specific downstream KILT dataset, and $i$ means that all relevant documents belong to 
$\{\mathcal{D}_0,\dots,\mathcal{D}_{i}\}$.

\begin{figure}[h]
  \centering 
  \vspace{-3mm}
  \includegraphics[width=0.65\linewidth]{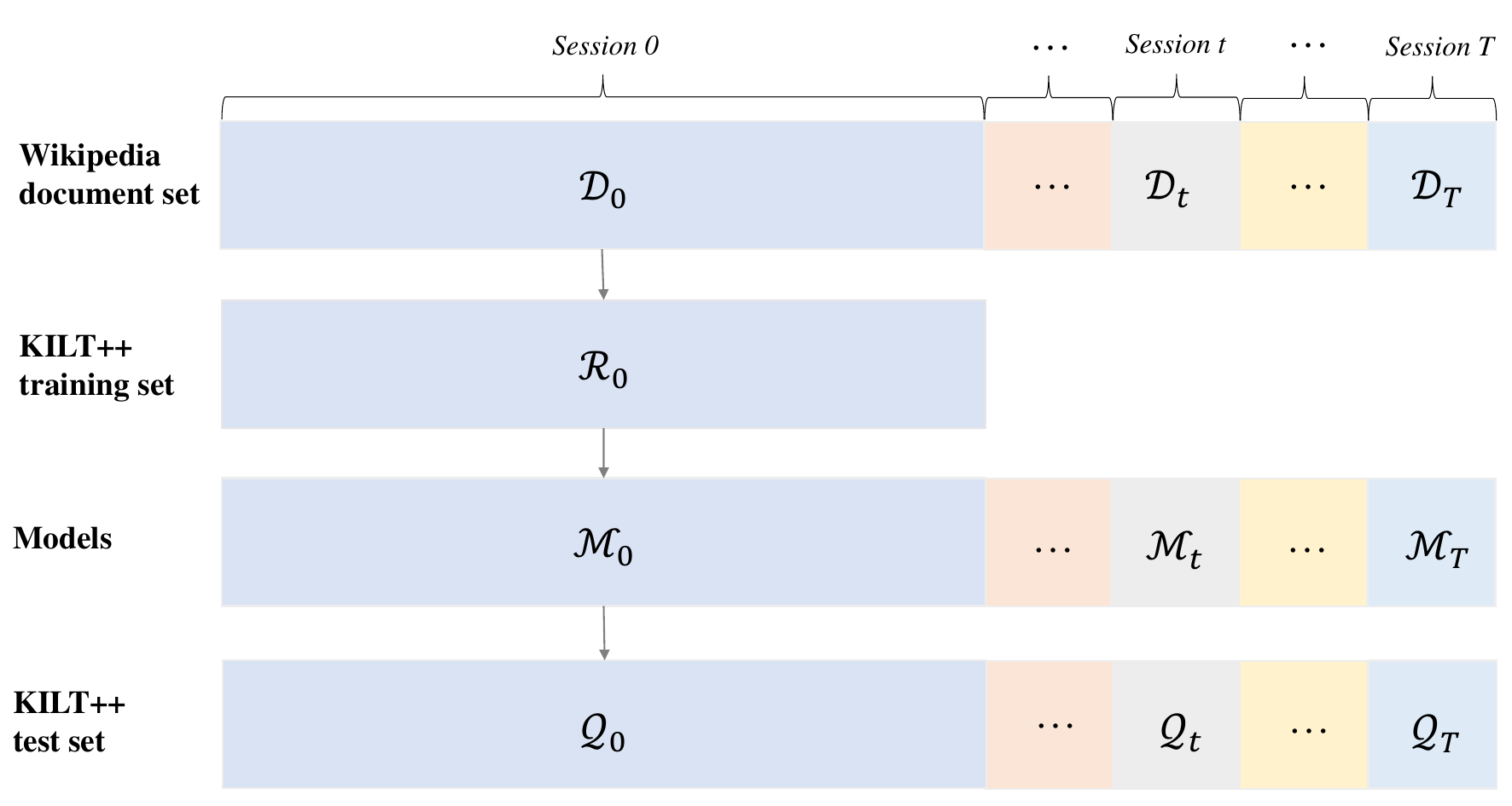}
  \caption{Evaluation criteria of the continual document learning task for KILTs.}
    \vspace{-3mm}
  \label{kilt++}
\end{figure}

\subsection{Benchmark construction}

In order to study and evaluate the CDL task for KILTs, we build a new benchmark dataset based on the original KILT dataset~\cite{petroni2020kilt}, i.e., KILT++. 
The KILT dataset encompasses eleven datasets spanning five knowledge-intensive language tasks, which are all rooted in a shared knowledge source derived from a common Wikipedia snapshot. 
We split the datasets in each task into $T+1$ sessions, i.e., session $0,\dots,T$, to simulate the dynamic retrieval scenario for KILTs.
We construct the benchmark dataset via the training set and the dev set since the KILT leaderboard imposes restrictions on the frequency of the submission for test performance. 

\begin{table}[t]
\caption{Overall statistics of our constructed KILT++ benchmark dataset.}
\label{statistics}
\centering
\setlength{\tabcolsep}{6pt}
\renewcommand{\arraystretch}{1}
\begin{tabular}{ll cccccc}
\toprule
\multirow{2}{*}{\textbf{Dataset}} & \multirow{2}{*}{\textbf{Task}} & \multicolumn{2}{c}{\textbf{0}}     & \textbf{1}      & \textbf{2}      & \textbf{3}      & \textbf{4}      \\ 
\cmidrule{3-4}
                                  &                                & \textbf{\#Train} & \textbf{\#Test} & \textbf{\#Test} & \textbf{\#Test} & \textbf{\#Test} & \textbf{\#Test} \\ \midrule
\textbf{FEV}                      & Fact Checking                  & \phantom{1,3}73,078            & 6,064            & 1,007            & 1,088            & 1,161            & 1,124            \\ 
\textbf{AY2}                      & Entity Linking                 & \phantom{1,3}10,745            & 2,847            & \phantom{1,}553             & \phantom{1,}418             & \phantom{1,}441             & \phantom{1,}525             \\ 
\textbf{WnWi}                     & Entity Linking                 & -                & 2,052            & \phantom{1,}360             & \phantom{1,}332             & \phantom{1,}288             & \phantom{1,}364             \\ 
\textbf{WnCw}                     & Entity Linking                 & -                & 3,381            & \phantom{1,}652             & \phantom{1,}460             & \phantom{1,}605             & \phantom{1,}501             \\ 
\textbf{T-REx}                    & Slot Filling                   & 1,304,897          & 2,876            & \phantom{1,}488             & \phantom{1,}504             & \phantom{1,}553             & \phantom{1,}579             \\ 
\textbf{zsRE}                     & Slot Filling                   & \phantom{1,3}94,980            & 2,226            & \phantom{1,}353             & \phantom{1,}379             & \phantom{1,}417             & \phantom{1,}349             \\ 
\textbf{NQ}                       & Open-domain QA                 & \phantom{1,3}57,238            & 1,353            & \phantom{1,}297             & \phantom{1,}359             & \phantom{1,}396             & \phantom{1,}432             \\ 
\textbf{HoPo}                     & Open-domain QA                 & \phantom{1,3}44,897            & 1,994            & \phantom{1,}742             & \phantom{1,}835             & \phantom{1,}959             & 1,070            \\ 
\textbf{TQA}                      & Open-domain QA                 & \phantom{1,3}33,152            & 2,457            & \phantom{1,}599             & \phantom{1,}732             & \phantom{1,}708             & \phantom{1,}863             \\ 
\textbf{ELI5}                     & Open-domain QA                 & -                & \phantom{1,}745             & \phantom{1,}172             & \phantom{1,}199             & \phantom{1,}201             & \phantom{1,}190             \\ 
\textbf{WoW}                      & Dialogue                       & \phantom{1,3}39,823            & 2,162            & \phantom{1,}391             & \phantom{1,}131             & \phantom{1,}199             & \phantom{1,}171             \\ \bottomrule
\end{tabular}
\vspace{-2mm}
\end{table}

To mimic the new arrival of documents, we set $T$ to $4$ and construct KILT++ as follows:
\begin{enumerate*}[label=(\roman*)]
    \item We randomly sample 60\% documents from the whole Wikipedia knowledge source to constitute the base document set $\mathcal{D}_0$. 
    Then, we randomly divide the remaining Wikipedia documents into four incremental sets with the same number of documents, to serve as $\mathcal{D}_1$, \ldots, $\mathcal{D}_4$.
    \item To construct the labeled query-document pairs $\mathcal{R}_0$ corresponding to $\mathcal{D}_0$, we filter the original KILT training set by retaining only those query-document pairs where all relevant articles in the corresponding provenance exclusively belong to $\mathcal{D}_0$.
    \item To construct the test sets $\mathcal{Q}_0$, \ldots, $\mathcal{Q}_4$ corresponding to $\mathcal{D}_0$, \ldots, $\mathcal{D}_4$, we employ an iterative algorithm as follows.
    Initially, we filter the original KILT dev set by retaining only those query-document pairs where all relevant articles in the corresponding provenance exclusively belong to $\mathcal{D}_0$ and denote the constructed dataset as $\mathcal{Q}_0$.
    As for constructing $\mathcal{Q}_i, i\ge1$, we iteratively filter the remaining KILT dev set by retaining only those query-document pairs where all relevant articles in the corresponding provenance exclusively belong to $\{\mathcal{D}_0,\dots,\mathcal{D}_i\}$ and denote the constructed dataset as $\mathcal{Q}_i$.
\end{enumerate*}

It is worth noting that since the original KILT dataset inherently consists of eleven datasets spanning five downstream tasks, the derived $\mathcal{R}_0$ and $\mathcal{Q}_{(.)}$ are essentially the same.
We typically use $\mathcal{R}_0$ or $\mathcal{Q}_{(\cdot)}$ to denote the KILT++ training or test set as a whole, and when we would like to elucidate a specific KILT dataset such as FEV, we can add superscripts as a differentiation, i.e., $\mathcal{R}_0^{FEV}$ or $\mathcal{Q}_{(\cdot)}^{FEV}$.
\autoref{statistics} shows the overall statistics of our KILT++ benchmark dataset. 

\subsection{Evaluation metrics}
In this section, we group the evaluation metrics into three parts. First, we describe the metrics employed for assessing individual downstream datasets. Subsequently, we elucidate the metrics utilized for assessing individual downstream tasks. Finally, we provide the metrics employed for assessing all downstream datasets.

\subsubsection{Assessing individual downstream datasets}
As illustrated in \autoref{statistics}, for each session $i$, we have eleven specific KILT++ test sets denoted as $Q_i^{\delta}$ where $\delta$ denotes the specific downstream dataset such as FEV and AY2, whose relevant documents belong to $\{\mathcal{D}_0,\dots,\mathcal{D}_i\}$.
Suppose the performance of the retrieval model $\mathcal{M}_t$ evaluated on the held-out test set $Q_i^{\delta}$ is $\mathcal{P}_{t,i}^{\delta}$,
\begin{equation} \label{performance}
    \mathcal{P}_{t,i}^{\delta}=\sum_{q \in Q_i^{\delta}, d_q\in\{\mathcal{D}_0,\dots,\mathcal{D}_i\}}g(d_q, \mathcal{M}_t(q)), i\leq t,
\end{equation}
where $d_q$ denotes the relevant document to the query $q \in Q_i^{\delta}$, and $g(\cdot)$ denotes a widely used evaluation metric for IR such as recall \cite{singhal2001modern}. 
Details of the evaluation metrics used will be described in Section \ref{metric introduction}. 

When it comes to individually assessing ${M}_t$ on a specific downstream dataset $\delta$, we compare the vertical performance $VP_t^{\delta}$ of different approaches on $Q_t^{\delta}$ in the same session $t$, 
\begin{equation}
    VP_t^{\delta}=\mathcal{P}_{t,t}^{\delta}.
\end{equation}

\subsubsection{Assessing individual downstream tasks}
When it comes to individually assessing ${M}_t$ on a specific downstream task $\tau$ in session $t$, we take the average vertical performance across all specific datasets under this task as the metric,
\begin{equation}
    VP_t^{\tau}=\frac{1}{\left|D^{\tau}\right|}\sum_{\delta\in D^{\tau}}VP_t^{\delta},
\end{equation}
where $D^{\tau}$ denotes the set of all specific downstream datasets that belong to the task $\tau$.

\subsubsection{Assessing all downstream datasets}
To give a comprehensive retrieval performance across all downstream datasets in the session $t$, we employ the vertical performance $VP_t$,
\begin{equation}
    VP_t=\frac{1}{\left|D\right|}\sum_{\delta\in D}VP_t^{\delta},
\end{equation}
where $D$ denotes the set of all specific downstream datasets.

Since we pay more attention to the comprehensive retrieval capability across all downstream datasets and tasks,
we merely employ the following across-all-session metrics to assess the trends in comprehensive retrieval performance.
Hence, we first define $\mathcal{P}_{t, i}$ to facilitate the elaboration of the later-defined metrics,
\begin{equation}
    \mathcal{P}_{t,i}=\frac{1}{\left|D\right|}\sum_{\delta\in D}\mathcal{P}_{t,i}^{\delta},
\end{equation}
where $D$ denotes the set of all specific downstream datasets.
To provide a metric for assessing all downstream datasets across all sessions, following \cite{lopez2017gradient, mehta2022dsi++}, we also employ the following evaluation metrics:
\begin{enumerate*}[label=(\roman*)]
    \item Average Performance ($AP$) to measure the average performance at the conclusion of training with the entire existing data sequence, 
    \item Backward Transfer ($BWT$) to evaluate the effect of learning a new session on the performance of all previous sessions, and 
    \item Forward Transfer ($FWT$) to measure the ability to learn when confronted with a new session, 
\end{enumerate*}
which are defined as follows:
\begin{equation}
    AP=\frac{1}{T+1}\sum_{i=0}^{T}\mathcal{P}_{T,i},
\end{equation}
\begin{equation}
    BWT=\frac{1}{T}\sum_{i=0}^{T-1}\max_{t\in\{0,\dots,T-1\}}(\mathcal{P}_{t,i}-\mathcal{P}_{T,i}),
\end{equation}
\begin{equation}
    FWT=\frac{1}{T}\sum_{t=1}^{T}\mathcal{P}_{t,t}.
\end{equation}

\section{Analysis Of Corpusbrain on CDL} 
\label{analysis}

In this section, based on our constructed KILT++ benchmark dataset, we conduct an empirical analysis of Corpusbrain to investigate its performance on CDL tasks. 

\subsection{Background} \label{background}

Our method is designed based on CorpusBrain~\cite{chen2022corpusbrain},  and thus we would like first to provide a brief overview of CorpusBrain before delving into the details of our proposed extension.
CorpusBrain is a pre-trained generative retrieval model for KILTs, exhibiting state-of-the-art retrieval performance for KILTs.

\subsubsection{Model architecture}
In CorpusBrain, a transformer-based~\cite{vaswani2017attention} encoder-decoder architecture is used to capture the relevance between queries and docids, which incorporates an encoder to yield the query representation and a decoder to generate the relevant docids.
In the implementation of CorpusBrain, the titles of Wikipedia pages are selected as docids.

\subsubsection{Pre-training tasks}

Three self-supervised pre-training tasks are devised to generate pseudo-query-docid pairs from documents and hence facilitate retrieval for KILTs. 
The pretraining tasks in CorpusBrain are carefully designed based on a prevailing hypothesis that utilizing pre-training tasks that bears greater resemblance to downstream tasks results in superior fine-tuning effectiveness.
Specifically, three pre-training tasks with different granularity are introduced:

\begin{itemize}[leftmargin=*]
    \item \textbf{Inner Sentence Selection (ISS).} Inner sentences are randomly sampled from the document as pseudo-queries, with the document and destination pages linked by anchor texts serving as relevant target documents. 
    The ISS task is designed to capture sentence-level semantic context.

    \item \textbf{Lead Paragraph Selection (LPS).} Leading paragraphs are drawn from the document as pseudo-queries, with document and destination pages linked by anchor texts serving as relevant target documents as well.
    The LPS task allows for capturing paragraph-level semantic information.

    \item \textbf{Hyperlink Identifier Prediction (HIP).} The corresponding sentences of randomly sampled anchors, along with the surrounding contextual sentences, are chosen as pseudo-queries, with the destination pages linked by the anchors serving as the relevant target documents. 
    The LPS task is leveraged to capture inter-document semantic relevance.
\end{itemize}

\subsubsection{Pre-training process}
\label{process}
In CorpusBrain, the ``pre-train and fine-tune'' paradigm is employed to adapt to multiple downstream KILT tasks.
In the pre-training phase, the checkpoint of BART \cite{lewis2020bart} is first applied to initialize the parameters to reduce the cost of training from scratch.
After generating pairs of pseudo-queries and docids by the aforementioned pretraining tasks, a standard Seq2seq learning objective, i.e., maximum likelihood estimation (MLE)~\cite{myung2003tutorial}, is employed to optimize the model, denoted as,
\begin{equation}
\mathcal{L}=\sum_{{q\in f(\mathcal{D})}}\sum_m\sum_n \log p(w_{m,n}\mid w_{\le m, <n},q;\Theta),
\end{equation}
where $\mathcal{D}$ represents the knowledge source corpus, $f(\cdot)$ signifies the transformation function of pretraining tasks, $q$ refers to the constructed pseudo-query, $w_{m,n}$ denotes the $n$-th token in the $m$-th docid related to $q$, and $\Theta$ refers to the model parameters.

\subsubsection{Fine-tuning process}
To further adapt to multiple downstream KILT tasks, the model is then fine-tuned on all KILT training datasets across five tasks through a multi-task training objective.
By applying the fine-tuned model to the KILT test set and decoding with a constrained beam search strategy on the docid prefix tree, as illustrated in the KILT leaderboard, CorpusBrain can achieve the top performance on a number of downstream tasks.

\begin{table}[t]
\caption{Performance of naive CorpusBrain variants on the CDL task. We evaluate the retrieval performance on $Q_i$ in terms of  $VP$. As for $AP$, $BWT$ and $FWT$, $\uparrow$ indicates higher is better and $\downarrow$ indicates lower is better.} 
\label{naive}
   \centering
    \setlength{\tabcolsep}{8pt}
    \renewcommand{\arraystretch}{1.1}
\begin{tabular}{l cccccccc}
\toprule
Model      & $\mathcal{Q}_0$ & $\mathcal{Q}_1$ & $\mathcal{Q}_2$ & $\mathcal{Q}_3$ & $\mathcal{Q}_4$ & \textbf{AP}$\uparrow$ & \textbf{BWT}$\downarrow$ & \textbf{FWT}$\uparrow$ \\ 
\midrule
BM25          &27.61 &25.11 &22.97 &23.84 &22.92 &22.26 &\phantom{0}2.78 &23.71 \\
Direct        &59.72 &47.86 &44.87 &46.21 &45.42 &48.28 &\phantom{0}0.67 &46.09 \\
Sequential   &59.72 &31.84 &27.09 &26.81 &22.48 &17.67 &19.89 &27.05    \\ \bottomrule
\vspace{-3mm}
\end{tabular}
\end{table}

\subsection{Overall performance} \label{naive analysis}
As for directly leveraging the pre-trained CorpusBrain model to solve the CDL task for KILTs, we build two naive CorpusBrain variants as follows.

\begin{itemize}[leftmargin=*]

\item \textbf{The direct insertion approach} (denoted as \textit{Direct}), inserts new docids, i.e., Wikipedia titles, from the incremental corpus $\mathcal{D}_t$ directly into the docid prefix tree, without updating the backbone parameter $\Theta_0$.

\item \textbf{The sequential pre-training approach} (denoted as \textit{Sequential}), sequentially pre-trains the model via $\mathcal{D}_0\sim\mathcal{D}_t$ with self-supervised pre-training tasks.

\end{itemize}

\heading{Experimental results} As illustrated in \autoref{naive}, the following observations can be made:
\begin{enumerate*}[label=(\roman*)]
\item The \textit{Direct} method suffers from a significant drop of approximately 20\% in terms of VP in the first incremental session, compared to BM25 with only 9\%, which demonstrates that the \textit{Direct} method cannot learn incremental documents well when they arrive.
Moreover, the \textit{Direct} method solely depends on the generalization capability of the backbone model, without a mechanism to learn knowledge within new documents.

\item The \textit{Sequential} variant exhibits a more considerable drop in terms of VP in the first incremental session, quantitatively about 47\%.
We can also observe a high BWT score, which demonstrates the \textit{Sequential} variant is prone to catastrophic forgetting.
Notably, the \textit{Sequential} method even demonstrates inferior performance compared to the traditional BM25 in terms of $VP$ in session 4.
\item On the whole, neither of the two naive variants can effectively tackle the CDL task for KILTs. 
\end{enumerate*}

Based on our analysis, we conclude that the naive variants of CorpusBrain are either deficient in their ability of forward transfer or prone to catastrophic forgetting, hence the CDL task for KILTs poses a non-trivial challenge for CorpusBrain.
\section{Methodology}

In this section, we introduce a novel continual generative pre-training framework for KILTs, i.e., CorpusBrain++.
We first introduce our model design, and then describe the technical details. 
Finally, we explain the learning and inference processes of the model.

\subsection{Model overview}
According to our analysis in Section \ref{analysis}, naive variants of CorpusBrain struggle to address the CDL challenge effectively and efficiently. 
We attribute this to the fact that the characteristics of continual learning for KILTs are neglected.
Since KILTs encompass multiple downstream tasks with distinct forms of input queries, modeling each task separately could facilitate continual learning. 

\begin{figure}[t]
  \centering 
  \includegraphics[width=\linewidth]{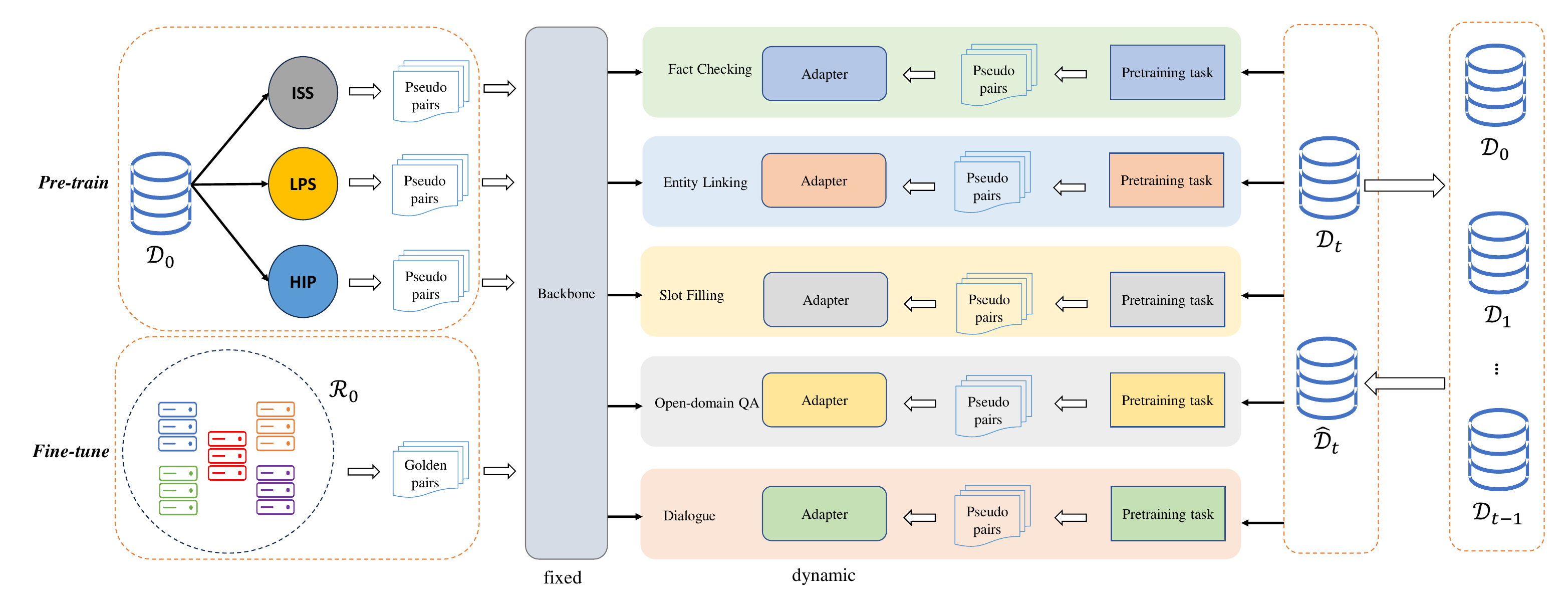}
  \caption{Illustration of our proposed CorpusBrain++ method. 
The backbone first involves pre-training on the initial base document set $D_0$ through the ISS, LPS, and HIP pretraining tasks, and fine-tuning with golden pairs derived from the KILT++ training set. 
To accommodate each task, a specific adapter is allocated, and a dedicated pretraining task is introduced to mimic the characteristics of downstream input queries. In addition to the incremental document set $D_t$, we also revisit semantically similar documents to $D_t$ from previous sessions, thereby generating pseudo pairs and continually pre-training the adapters.}
\vspace{-3mm}
  \label{model}
\end{figure}

Inspired by the dynamic-architecture paradigm in the continual learning field \cite{mallya2018packnet, rusu2016progressive}, we propose a continual generative pre-training framework to incrementally learn new documents for KILTs, namely CorpusBrain++. 
As shown in \autoref{model}, CorpusBrain++ incorporates three key features: 
\begin{enumerate}[label=(\arabic*),leftmargin=*]
    \item First, we employ a backbone-adapter architecture. The shared backbone undergoes pre-training on $\mathcal{D}_0$ and subsequent fine-tuning on $\mathcal{R}_0$, with the parameters held fixed to ensure consistent provision of task-shared knowledge. To accommodate the evolving corpus, a specific adapter is used for each KILT task to efficiently learn new documents, which aims to retain the characteristics of specific data in each task.
    \item Second, we design a specific pre-training objective for each KILT task, to resemble the relevant relationship between queries and documents in each specific task. In this way, we can continually pre-train the task-specific adapters. 
    \item Third, to avoid catastrophic forgetting, we revisit some old documents and apply the specific pre-training tasks for both new and old documents for continual pre-training.
\end{enumerate}

\subsection{Shared backbone}
\begin{figure}[t]
  \centering 
  \includegraphics[width=\linewidth]{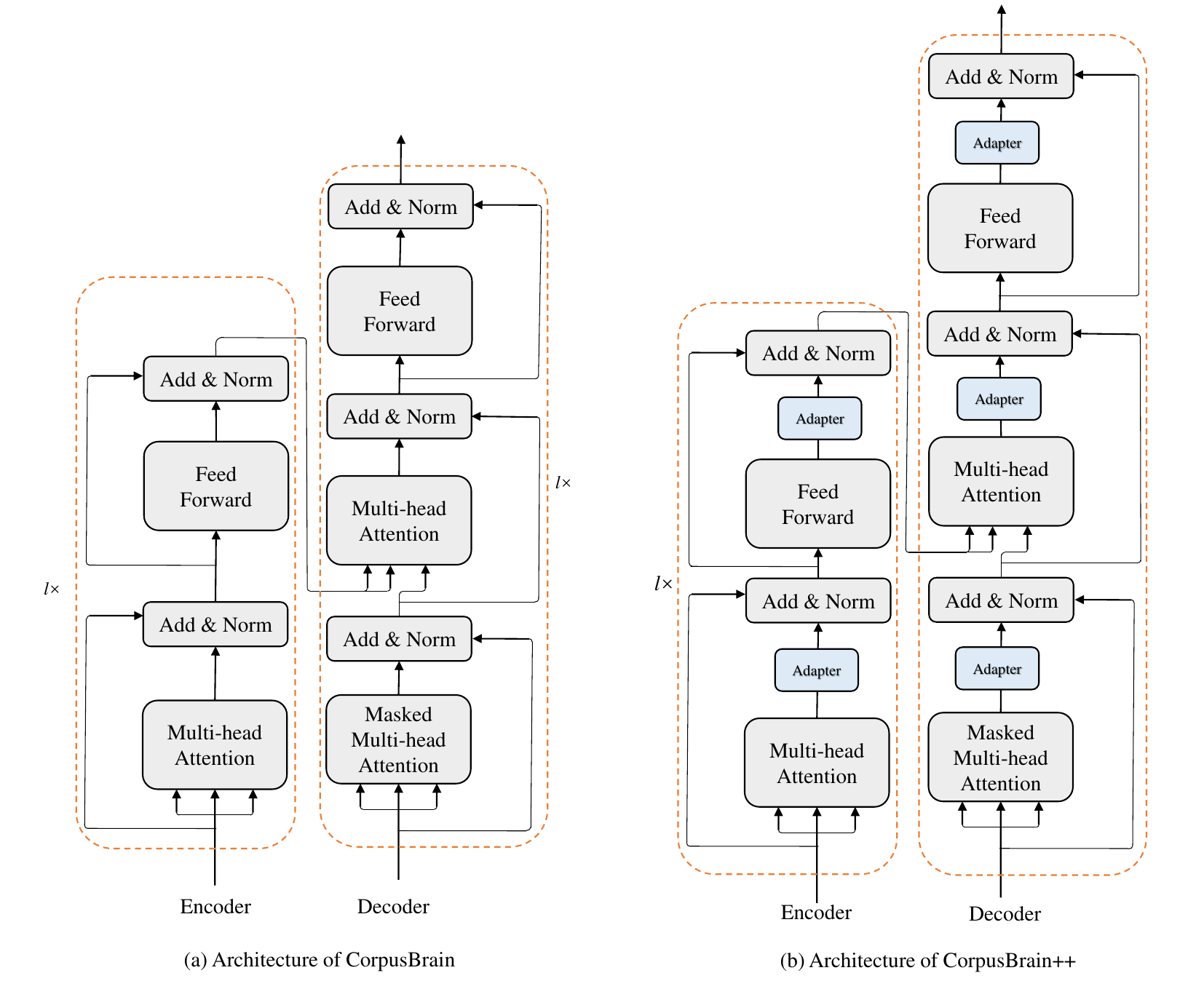}
  \caption{The architecture of CorpusBrain and CorpusBrain++.}
  \label{insertion}
\end{figure}

Here, we present the structure of the shared backbone, which constitutes a transformer-based encoder-decoder architecture.
As depicted in \autoref{insertion} (a), both the encoder and decoder are generally composed of $l$ layers.
Each layer of the encoder comprises multiple components including a multi-head self-attention sub-layer ($MHSA$), a feed-forward neural network sub-layer ($FFN$), and a residual connection subsequently followed by layer normalization ($RCLN$).
Likewise, each layer of the decoder consists of multiple components incorporating a masked multi-head self-attention sub-layer ($MMHSA$), a multi-head cross-attention sub-layer ($MHCA$), a feed-forward neural network sub-layer ($FFN$), and a residual connection subsequently followed by layer normalization ($RCLN$).

Next, we introduce the basic multi-head attention mechanism, then specify three sub-layers incorporating the attention mechanism and introduce the principle of the feed-forward layer.

\heading{Multi-head Attention Mechanism} Given the input hidden state $h^q, h^k, h^v \in \mathbb{R}^{n \times d}$ , the $i$-th attention head can be formulated as:
\begin{equation}
\operatorname{Attention}_i(\textbf{h}^q,\textbf{h}^k,\textbf{h}^v)=\sum_{m}\operatorname{softmax}\left(\frac{W_i^q \textbf{h}^q\cdot W_i^k \textbf{h}^k}{\sqrt{d/m}}\right)W_i^v\textbf{h}^v,
\end{equation}
where $W_i^{(\cdot)} \in \mathbb{R}^{d/m \times m}$ are trainable projection matrices. 
Additionally, 
the outputs of multiple attention heads are fed into a 
multi-head attention layer, of which the mechanism can be formulated as follows:
\begin{equation}
    MH(\textbf{h}^q, \textbf{h}^k, \textbf{h}^v) = \operatorname{Concat}(\operatorname{Attention}_1(\textbf{h}^q,\textbf{h}^k,\textbf{h}^v),\dots,\operatorname{Attention}_n(\textbf{h}^q,\textbf{h}^k,\textbf{h}^v))W^o,
\end{equation}
where $W^o \in \mathbb{R}^{d \times d}$ is the learned transformation matrix.

In the self-attention layers, i.e., $MHSA$ and $MMHSA$, $\textbf{h}^q$, $\textbf{h}^k$ and $\textbf{h}^v$ all refer to the input hidden state $\textbf{h}$. 
An attention mask is employed in $MMHSA$ to preserve the auto-regressive property.
In the cross-attention layers, i.e., $MHCA$, $\textbf{h}^q$ comes from the previous decoder layer while $\textbf{h}^k$ and $\textbf{h}^v$ come from the output of the encoder layer.

\heading{Feed-forward Sub-layer} The feed-forward sub-layer is a position-wise fully connected FFN, which can be formulated as follows:
\begin{equation}
FFN(\textbf{h})=\sigma(\textbf{h} W_1+b_1)W_2+b_2,
\end{equation}
where $W_1 \in \mathbb{R}^{d\times 4d}$ and $W_2 \in \mathbb{R}^{4d\times d}$ are trainable transformation matrices, $b_1 \in \mathbb{R}^{4d}$ and $b_2\in \mathbb{R}^{d}$ are trainable bias terms.

Each sub-layer, denoted as $S\in \{MHSA, MMHSA, MHCA, FFN\}$, uses a residual connection followed by layer normalization ($RCLN$), which can be formulated as follows:
\begin{equation} \label{sublayer}
RCLN(\textbf{h})=LN(S(\textbf{h})+\textbf{h}),
\end{equation}
where $LN(\cdot)$ denotes layer normalization.

\subsection{Task-specific adapter}
As mentioned before, we assign a task-specific adapter for each downstream KILT task and continually pre-train the adapters to accomplish the CDL task for KILTs. 
Below, we provide the technical details for adapter insertion and adapter structure.

\subsubsection{Adapter insertion}
To capture a specific query-to-docid mapping for each task, we maintain an adapter module for each task independently, i.e., the task-specific adapter. 
The adapter is a series of compact and efficient modules inserted after the sub-layers of the transformer, defined as, 
\begin{equation}
\label{adapter}
\operatorname{Adapter}(\textbf{h})=g(\textbf{h}W_{down})W_{up},
\end{equation}
where $W_{down}\in \mathbb{R}^{d\times r}$ and $W_{up}\in \mathbb{R}^{r\times d}$ are trainable down-projection and up-projection matrices correspondingly, and $g$ denotes the transformation function.

As depicted in \autoref{insertion}(b), we insert adapters into both the encoder and decoder modules.
The underlying reasons are two-fold:
\begin{enumerate*}[label=(\roman*)]
    \item The input queries vary significantly across distinct downstream tasks, including semantic granularity and formats, hence we insert adapters into the encoder module.
    \item The query-docid mapping also varies across downstream tasks, therefore we insert adapters into the decoder module apart from the encoder module.
\end{enumerate*}

Given the promising results achieved in \cite{houlsby2019parameter, su2020continual}, as depicted in \autoref{insertion}(b), we insert adapters in an inside way.
In other words, we place an adapter behind each sub-layer $S\in \{MHSA$, $MMHSA$, $MHCA$, $FFN\}$.
The formulation of each sub-layer in Eq.~\ref{sublayer} undergoes the following transformation,
\begin{equation}
       \operatorname{Inside}(\textbf{h})=\operatorname{LN}(\operatorname{Adapter}(S(\textbf{h}))+\textbf{h}).
\end{equation}

\subsubsection{Adapter structure}
In the area of multi-task learning and transfer learning~\cite{houlsby2019parameter}, previous work has made much progress on the issue of how to design more efficient transformation functions $g$ in  Eq.~\ref{adapter}.  
We employ a simple yet effective adapter structure named the low-rank layer, which exhibits promising empirical results in \cite{stickland2019bert, chen2021fedmatch}.
In this adapter structure, the $g$ function is defined as a low-rank transformation, i.e., an identity function.

\subsection{Task-specific pre-training objective} \label{pre-training objective}
To facilitate continually pre-training the task-specific adapters, we carefully design a specific pre-training objective for each downstream KILT task.
The principle of each pre-training task is to mimic the relevant relationship between queries and documents in the corresponding downstream task as much as possible.
The objective of each pre-training task mainly incorporates two components, i.e., input pseudo-queries and output docids.
In terms of distinct downstream KILT tasks, distinctions of input pseudo-queries primarily manifest themselves in semantic granularity.
During the construction of input pseudo-queries, we first build preliminary pseudo-queries based on semantic granularity, and further refine them to accommodate the characteristics of downstream KILT tasks.
On the other hand, distinctions of output docids primarily manifest themselves in the number of supporting documents.
During the construction of output docids, we determine the number of output docids for each downstream task according to the specific output scenario associated with the corresponding task.

\subsubsection{Fact checking}
\begin{figure}[t]
  \centering 
  \includegraphics[width=\linewidth]{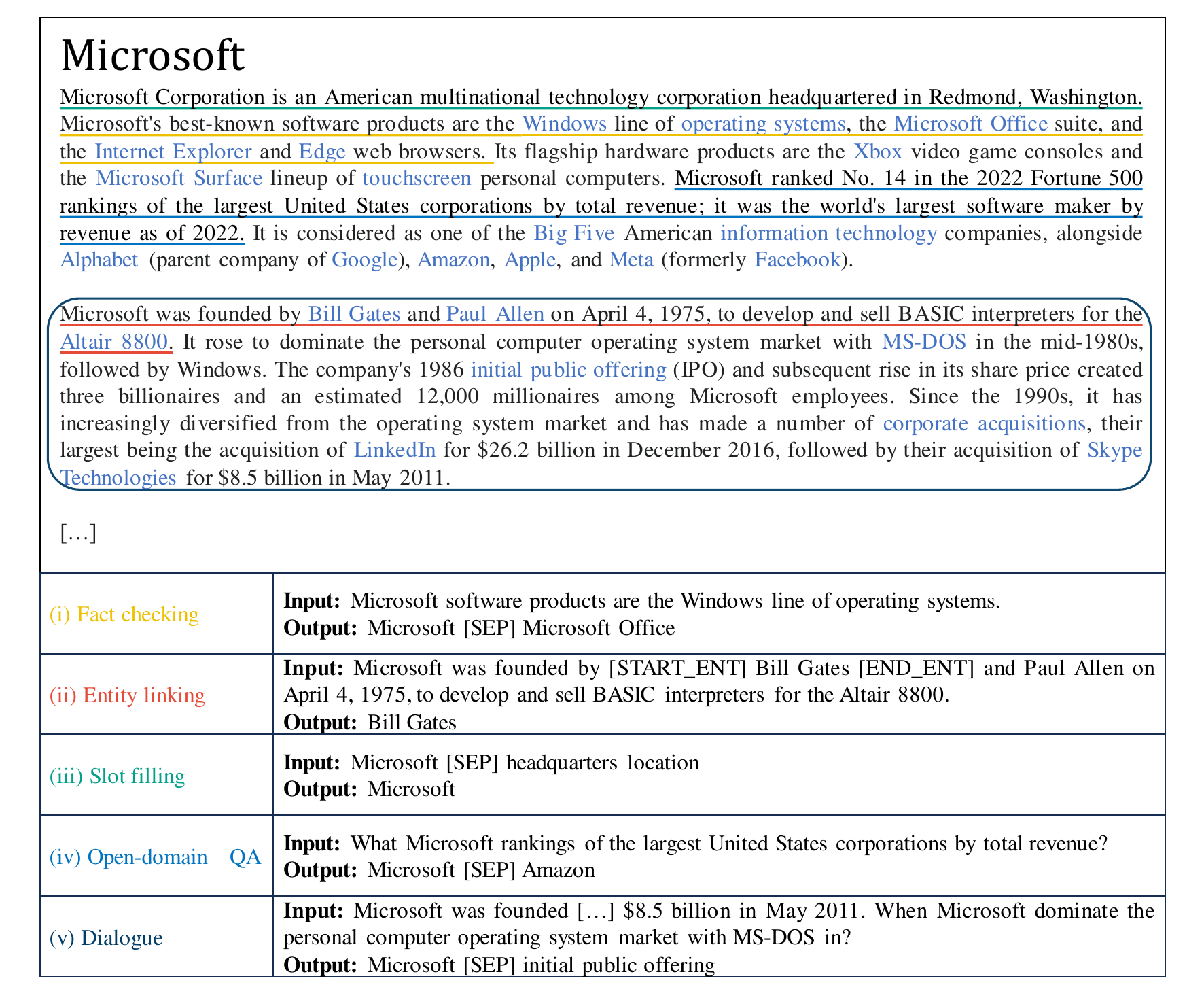}
  \caption{Illustration of specific pre-training tasks for each KILT task. Anchor texts are marked in blue. The colored underlines in Wikipedia content correspond to the source text of the corresponding KILT task in the table below. Query examples provide an example of the corresponding downstream KILT task. Input and output refer to the constructed input pseudo-queries and corresponding output docids.} 
  \label{pre-train}
  \vspace{-2mm}
\end{figure}

In terms of fact checking, the input pseudo-queries and output docids are constructed as follows:
\begin{itemize}[leftmargin=*]
    \item \heading{Input pseudo-queries} The construction of input pseudo-queries includes the following steps:
    \begin{enumerate}[label=(\arabic*),leftmargin=*]
    \item     From the perspective of semantic granularity, fact-checking queries typically encompass sentence-level semantic context.
    For instance, a typical query example that \emph{Windows are the software products of Microsoft} is at the sentence level.
    Hence we use the ISS pre-training task to sample sentences from the document.
    Specifically, given a document $d$, we randomly draw $l$ inner sentences from $d$ to form the preliminary pseudo-queries.
    \item     We can also observe that fact-checking queries are typically short in length (7 words in this case) and entity-centric (\emph{Windows} and \emph{Microsoft} in this case).
    To further mimic the aforementioned characteristics, we randomly sample an $n$-gram span from each preliminary pseudo-query.
    \item     The document title is regarded as the core entity within $d$ in previous work~\cite{de2020autoregressive, wu2020scalable}.
    Therefore, We add the document title to the beginning of the sampled span to construct the final pseudo-query. An example input pseudo-query for fact checking is given in \autoref{pre-train}(i).
    \end{enumerate}

    \item \heading{Output docids} 
    From the perspective of target document numbers, fact checking might require multiple supporting documents to judge the authenticity of a given claim.
    For example, we might need two supporting documents in this case, i.e., a document titled \emph{Windows} and another titled \emph{Microsoft}.
    Following ISS, we first randomly sample $o$ anchor texts within $d$.
    Apart from $d$, we also treat the destination pages linked by these $o$ anchors as the relevant documents.
    As depicted in \autoref{pre-train}(i), we concatenate the docid of $d$ and the docids of these $o$ relevant documents with a separator \texttt{[SEP]}.
    By this means the final output sequence could be constructed, which allows for dynamic predictions of relevant documents.
\end{itemize}

\subsubsection{Entity linking}
In terms of entity linking, the input pseudo-queries and output docids are constructed as follows:
\begin{itemize}[leftmargin=*]
    \item \heading{Input pseudo-queries} The construction of input pseudo-queries includes the following steps:
    \begin{enumerate}[label=(\arabic*),leftmargin=*]
    \item     From the perspective of semantic granularity, entity-linking queries typically involve inter-document semantic relations.
    We provide a typical entity-linking query as an example: \texttt{[START\_ENT]} \emph{Bill Gates} \texttt{[END\_ENT]}  \emph{is best known as the co-founder of Microsoft Corporation, one of the world's largest and most successful technology companies.}
    This query carries the inter-document semantic relation between the document titled \emph{Bill Gates} and the document titled \emph{Microsoft}.
    Based on the above analysis, we employ a variant of the HIP pre-training task to construct preliminary pseudo-queries.
    Specifically, given a document $d$, we randomly select $l$ anchor texts within $d$.
    Subsequently, we locate the corresponding sentences containing the selected anchor texts as the preliminary pseudo-queries.
    \item  As illustrated in \autoref{pre-train}(ii), entity-linking queries typically contain special tokens to indicate the entity boundary, i.e., \texttt{[START\_ENT]} and \texttt{[END\_ENT]}.
    To further mimic this characteristic, we insert special tokens revealing entity boundaries as well.
    Specifically, for each preliminary pseudo-query, we insert \texttt{[START\_ENT]} and \texttt{[END\_ENT]} to the left and right of the anchor text. An example input pseudo-query for entity linking is given in \autoref{pre-train}(ii).
    \end{enumerate}

    \item \heading{Output docids}
    From the perspective of target document numbers, only a unique target docid is required for entity linking (\emph{Bill Gates} in this case).
    To this end, for each pseudo-query, the output docid refers to the docid of the destination page linked by the selected anchor text.
    As illustrated in \autoref{pre-train}(ii), the constructed output docid refers to \emph{Bill Gates} in this case.
\end{itemize}

\subsubsection{Slot filling}
In terms of slot filling, the input pseudo-queries and output docids are constructed as follows:
\begin{itemize}[leftmargin=*]
    \item \heading{Input pseudo-queries}  The construction of input pseudo-queries includes the following steps:
    \begin{enumerate}[label=(\arabic*),leftmargin=*]
    \item  From the perspective of semantic granularity, slot-filling queries typically involve sentence-level semantic context.
    For example, a typical slot-filling query \emph{Microsoft} \texttt{[SEP]} \emph{headquarters location} mainly contains sentence-level semantics.
    Therefore, we first use the ISS pre-training task to sample sentences from the document. 
    Given a document $d$, we randomly draw $l$ inner sentences from $d$ to form the preliminary pseudo-queries.  
    \item    Each slot-filling query typically constitutes a subject entity (\emph{Microsoft} in this case) and a relational predicate (\emph{headquarters location} in this case) pre-defined in the candidate set.
    To closely mimic this characteristic, we first train a relation detector by fine-tuning the BERT model~\cite{devlin2019bert} on the slot-filling training set in the initial session, i.e., $\mathcal{R}_0^{T-REx}$ and $\mathcal{R}_0^{zsRE}$.
    The relation detector allows for the prediction of relation types within each sampled sentence. 
    We retain the top-$k$ predicted relation types for each sampled sentence.
    \item    We treat the core entity within $d$, i.e., the document title, as the subject entity.
    Subsequently, for each sampled sentence, we separately concatenate the subject entity with the top-$k$ predicted relation types using a separator (\texttt{[SEP]}) to form $k$ final pseudo-queries.
    An example input pseudo-query for slot filling is given in \autoref{pre-train}(iii).
    \end{enumerate}
   
    \item \heading{Output docids}
    From the perspective of target document numbers, the slot-filling task tends to require a single supporting document (\emph{Microsoft} in this case).
    Therefore, as depicted in \autoref{pre-train}(iii), the output docid refers to the docid of $d$ for each pseudo-query.
\end{itemize}

\subsubsection{Open-domain QA}
In terms of open-domain QA, the input pseudo-queries and output docids are constructed as follows:
\begin{itemize}[leftmargin=*]
    \item \heading{Input pseudo-queries}  The construction of input pseudo-queries includes the following steps:
    \begin{enumerate}[label=(\arabic*),leftmargin=*]
    \item  In the case of open-domain QA, queries typically involve sentence-level semantic context.
    Therefore, we first leverage the ISS pre-training task to sample sentences from the document. 
    Given a document $d$, we randomly draw $l$ inner sentences from $d$ to form the preliminary pseudo-queries.
    \item As illustrated in \autoref{pre-train}(iv), queries for open-domain QA are typically in the form of questions.
    Additionally, we can observe that QA queries are often short in length (10 words in this case).
    Moreover, the queries could sometimes incorporate an entity to locate the supporting documents (\emph{Microsoft} in this example).
    To further mimic the aforementioned characteristics, for each preliminary pseudo-query, we randomly sample an $n$-gram span.
    Subsequently, for each sampled span, we randomly select an interrogative word from a pre-defined candidate set, and then add it to the beginning of the span.
    \item We insert the document title between the interrogative word and the sampled span to emphasize the core entity within $d$.     
    An example input pseudo-query for open-domain QA is given in \autoref{pre-train}(iv).
    \end{enumerate}
   
    \item \heading{Output docids}
    From the perspective of target document numbers, more than one supporting document might be required to accomplish the QA task.
    Following ISS, we first randomly sample $o$ anchor texts within $d$.
    Subsequently, we treat the destination pages linked by these $o$ anchors as the relevant documents.
    Finally, as shown in \autoref{pre-train}(iv), we concatenate the docid of $d$ and the docids of the $o$ relevant documents with a separator \texttt{[SEP]}.
    By this means the final output sequence could be constructed, which allows for dynamic predictions of relevant documents.
\end{itemize}

\subsubsection{Dialogue}
In terms of dialogue, the input pseudo-queries and output docids are constructed as follows:
\begin{itemize}[leftmargin=*]
    \item \heading{Input pseudo-queries} The construction of input pseudo-queries includes the following steps:
    \begin{enumerate}[label=(\arabic*),leftmargin=*]
    \item  From the perspective of semantic granularity, queries in the dialogue typically involve paragraph-level semantic information.
    Hence, we first apply the LPS pre-training task to sample paragraphs from the document. 
    Given a document $d$, we sample the leading $l$ paragraphs from $d$ to form the preliminary pseudo-queries.
    \item  As illustrated in \autoref{pre-train}(v), dialogue queries tend to comprise a long conversation context and a question related to the context.
    To further mimic this characteristic, we treat each preliminary pseudo-query as a conversation context and construct a question for each preliminary pseudo-query.
    For each preliminary pseudo-query, we randomly sample an $n$-gram span.     
    Subsequently, for each sampled span, we randomly select an interrogative word from a pre-defined candidate set, and then add it to the beginning of the span. We insert the document title between the interrogative word and the sampled span to emphasize the core entity within $d$. By this means we could construct $l$ paragraph-level questions.
    \item  Instead of replacing the preliminary pseudo-queries with the constructed questions, we concatenate the preliminary pseudo-queries and the constructed questions to yield the final pseudo-queries.     An example input pseudo-query for dialogue is given in \autoref{pre-train}(v).
    \end{enumerate}

    \item \heading{Output docids}
    From the perspective of target document numbers, multiple supporting documents might be needed to comprehend the conversation context and provide a correct answer.
    Following LPS, we first randomly sample $o$ anchor texts within $d$.
    Subsequently, we treat the destination pages linked by these $o$ anchors as the relevant documents.
    As shown in \autoref{pre-train}(v), we concatenate the docid of $d$ and the docids of the $o$ relevant documents with a separator \texttt{[SEP]}.
    By this means the final output sequence could be constructed, which allows for dynamic predictions of relevant documents.
\end{itemize}

\subsection{Learning process}
Here, we detail the learning process deployed in our method. First, we elucidate the learning process of newly arrived documents. We then explain the strategy of rehearsing old documents, which serves as a countermeasure against catastrophic forgetting. Lastly, we present the overall learning objective. 

\subsubsection{Learning new documents} 
\label{new}
In the learning process, we continually learn new documents by updating the parameters of task-specific adapters.
When confronted with a new document set $D_t, t\ge1$, we first generate pairs of pseudo-queries and docids by applying the task-specific pre-training tasks to each incremental document.
For each downstream task, we can learn new documents by continually pre-training the corresponding task-specific adapter with the constructed pairs.
The learning objective is detailed in Section \ref{learn}.

\subsubsection{Rehearsing old documents} \label{old}
To svoid catastrophic forgetting of old documents while learning new documents, we perform experience rehearsal of old documents.
When learning new knowledge or skills, humans tend to draw upon similar past experiences as a reliance~\cite{haviland1974s, clark1974psychological, clark1977discourse}.
Inspired by this fact about human cognition, we employ a cluster-based strategy for rehearsing old documents in our framework, which allows for the consideration of semantic similarity.
For each new session $t\ge 1$, we first cluster old documents in previous sessions into $k$ categories.
In our framework, we employ the K-means clustering algorithm~\cite{hamerly2003learning} to cluster documents in $\mathcal{D}_0,\dots,\mathcal{D}_{t-1}$ into $k$ categories.
In terms of representing each document, we directly leverage the docid to facilitate efficiency.
We feed all docids of the old documents into the K-means algorithm,
\begin{equation}
    \{C_1,\dots,C_k\}=\operatorname{K\text{-}means}(\{\gamma \mid \gamma\in\bigcup_{i=0}^{t-1} \mathcal{D}_i\}),
\end{equation}
where $\gamma$ denotes the docid, $C_{(\cdot)}$ refers to the document cluster.

When a stream of documents $\mathcal{D}_t$ arrives, for each new document $d_t^i\in \mathcal{D}_t$, we first judge the specific cluster to which the document belongs.
Subsequently, we randomly select $n$ old documents from the corresponding cluster.
After repeating the aforementioned process for each $d_t^i\in \mathcal{D}_t$, we can construct an experience set $\hat{\mathcal{D}}_t$, which share semantic similarity with $\mathcal{D}_t$.
Similar to learning new documents, we leverage the aforementioned pre-training tasks on $\hat{\mathcal{D}}_t$ to generate pairs of pseudo-queries and docids.
For each downstream task, we can review old documents by continually pre-training the corresponding task-specific adapter with the constructed pairs.
The learning objective is detailed in Section \ref{learn}.

\subsubsection{Overall learning objective}
\label{learn}
For each new session $t$, the parameters of each task-specific adapter are independently updated, while the parameters of the shared backbone are kept unchanged. 
First, we initialize the parameters of each adapter by inheriting from the previous session.
Under exceptional circumstances wherein $t=1$, the parameters of task-specific adapters are randomly initialized since it's the first incremental session.
Subsequently, we construct pairs of pseudo-queries and docids for both the new document set $\mathcal{D}_t$ and the old document set $\hat{\mathcal{D}}_t$ following Section \ref{new} and Section \ref{old}.
Finally, for each downstream KILT task, we separately update the parameters of the task-specific adapter with the generated pairs tailored for the corresponding task.
We apply a standard sequence-to-sequence learning objective to continually pre-train the task-specific adapters,
\begin{equation}
\mathcal{L}=\sum_{\tau\in\mathcal{T}}\sum_{{(q,\gamma)\in f^{\tau}(\mathcal{D}_t\cup\hat{\mathcal{D}}_t)}}\log p(\gamma,\Theta^{\tau}_t\mid q;\Theta^{\tau}_{t-1}),
\end{equation}
where $\mathcal{T}$ denotes the task set, $\tau$ represents a specific downstream task in $\mathcal{T}$, $f^{\tau}(\cdot)$ refers to the transformation function of the pre-training task dedicated for $\tau$, $(q,\gamma)$ denotes the constructed pairs of pseudo-queries and docids. 
$\mathcal{D}_t$ refers to the new document set, and $\hat{\mathcal{D}}_t$ refers to the old document set derived from the cluster-based strategy of document rehearsal.
$\Theta^{\tau}_{t-1}$ represents the meta parameters of the task-specific adapter dedicated for $\tau$ before the $t$-th update, and  $\Theta^{\tau}_{t}$ refers to the meta parameters of the task-specific adapter dedicated for $\tau$ after the $t$-th update.

\subsection{Inference process}
During the inference phase, when confronted with the test set $Q_t,t\ge1$,
we initially categorize the test data into distinct task groups. Subsequently, for the test data of each task, we activate both the backbone model and the corresponding task-specific adapter to generate target docids incorporating both task-shared and task-specific knowledge. Additionally, we use a constrained beam search approach~\cite{anderson2017guided, de2020autoregressive} to confine the generated docids within a predefined set of docids.
It is noteworthy that, in session $t$, only docids corresponding to existing documents, namely, $\mathcal{D}_0,\dots,\mathcal{D}_t$, are incorporated into the prefix tree.
Since docids refer to document titles in this work, we confine the output sequence within the constraints of a document title prefix tree. 
Specifically, each node in the prefix tree corresponds to a token. 
When traversing from the root to a specific node, all nodes on the path collectively constitute a document title.

\section{EXPERIMENTAL SETTINGS}
In this section, we explain our experimental settings.

\subsection{Models}
\subsubsection{Baselines}
In this study, we conduct a comparative analysis involving our proposed CorpusBrain++, traditional IR models, and generative IR models.

\begin{itemize}[leftmargin=*]
\item \heading{Traditional IR models}
\begin{enumerate*}[label=(\roman*)]
    \item \textbf{BM25} is a typical sparse retrieval model, which utilizes term-based features to model the relevance between queries and documents.
    \item \textbf{DPR} is a representative dense retrieval model, which models the semantic relevance between queries and documents via a dual-encoder architecture.
\end{enumerate*}
For the traditional IR models, our empirical results encompass both incremental and non-incremental scenarios thanks to the high reproducibility provided by Pyserini\footnote{\url{https://github.com/castorini/pyserini}} and Tevatron.\footnote{\url{https://github.com/texttron/tevatron}}

\item \heading{Generative IR models}
We first consider several generative IR models in stationary scenarios, including
\begin{enumerate*}[label=(\roman*)]
    \item \textbf{GENRE}~\cite{de2020autoregressive}, which directly fine-tunes BART via multi-task training on the labeled KILT training datasets and supervised BLINK datasets\cite{wu2020scalable}; and
    \item \textbf{SEAL}~\cite{bevilacqua2022autoregressive}, which applies a BART-based autoregressive search engine to generate distinctive n-grams as docids.
\end{enumerate*}
Given that the focus of both models is confined to the non-incremental scenario, we only compare them with CorpusBrain++ in terms of non-incremental retrieval performance.
In the non-incremental scenario, CorpusBrain++ degenerates to CorpusBrain, and the experimental contrast between CorpusBrain and these two models has been previously conducted and reported in \cite{chen2022corpusbrain}. Furthermore, we explore some advanced generative IR models fitting in dynamic scenarios, including
\begin{enumerate*}[label=(\roman*)]
    \item \textbf{DSI++}, which continually fine-tunes DSI over new documents and allocates a unique integer as the docid for each new document; and 
    \item \textbf{CLEVER}, which introduces a technique named incremental product quantization to assign a docid to each new document.
\end{enumerate*}
\end{itemize}

\subsubsection{Ablation models}
Apart from the naive \textbf{Direct} and \textbf{Sequential} variants, whose modifications on top of CorpusBrain have been introduced in Section \ref{naive}, we modify CorpusBrain++ from three perspectives to explore the effectiveness of each component:

\begin{itemize}[leftmargin=*]
    \item \heading{The impact of different model architectures} We assess two variants of CorpusBrain++, which continually pre-train the backbone rather than the adapter with the task-specific pre-training objective:
    \begin{enumerate*}[label=(\roman*)]
        \item For \textbf{$\text{CorpusBrain++}_{-Adapter(ST)}$} we continually pre-train the backbone in a single-task manner, i.e., we continually pre-train the backbone independently for each downstream task with the corresponding task-specific pretraining objective. And
        \item for \textbf{$\text{CorpusBrain++}_{-Adapter(MT)}$} we continually pre-train the backbone in a multi-task manner, i.e., we continually pre-train the backbone jointly for all downstream tasks with the proposed task-specific pretraining objective.
    \end{enumerate*}

    \item \heading{The impact of different pre-training objectives} We design a variant
    \textbf{$\text{CorpusBrain++}_{OriPT}$}, where we continually pre-train the task-specific adapter with the original ISS, LPS, and HIP pretraining tasks.

    \item \heading{Analyze the impact of different document rehearsal strategies} We implement two variants:
    \begin{enumerate*}[label=(\roman*)]
        \item \textbf{$\text{CorpusBrain++}_{Random}$}, where we randomly sample some old documents to construct query-docid pairs via the task-specific objective; and
        \item \textbf{$\text{CorpusBrain++}_{-Rehearsal}$}, where we eliminate the strategy of old document rehearsal, i.e., we solely construct query-docid pairs with incremental documents.
    \end{enumerate*}
\end{itemize}

\subsection{Implementation details}
In this work, we utilize the Wikipedia document title as the docid for simplicity.
Following \citet{chen2022corpusbrain}, we employ $BART_{large}$ as the backbone architecture.
We can divide the learning process into two phases, i.e., the initial phase and the incremental phase.

In the initial phase, we incorporate the pre-training and fine-tuning process following \citet{chen2022corpusbrain}.
For the pre-training process, we utilize three pre-training tasks originally defined in \cite{chen2022corpusbrain}, i.e., ISS, LPS, and HIP, to construct pre-training data on $\mathcal{D}_0$, and the number of constructed pseudo-pairs is determined the same as \cite{chen2022corpusbrain}. 
Additionally, we employ a learning rate of $3e^{-5}$ alongside the Adam optimizer\cite{kingma2014adam}, incorporating a warmup technique with a warmup ratio of 0.1.
Moreover, we set the weight decay to 0.01, set the label smoothing to 0.1, set the gradient norm clipping to 0.1, and set the batch size to 8192 tokens.
In terms of the fine-tuning process, we fine-tune the backbone model via multi-task training on $\mathcal{R}_0$ spanning five distinct downstream tasks following \cite{bevilacqua2022autoregressive,de2020autoregressive,chen2022corpusbrain}, it is worth noting, as illustrated in \autoref{statistics}, that not all eleven datasets in $\mathcal{R}_0$ have an accessible training set.
We set the learning rate to $3e^{-5}$ and set the batch size to 4096 tokens.

In the incremental phase, i,e, in session $t, t\ge 1$, we first construct pre-training data by the well-designed task-specific pre-training tasks
with the newly arrived documents $\mathcal{D}_{t}$ and the revisited documents $\hat{\mathcal{D}}_{t}$, and then continually pre-train the task-specific adapters via the constructed pre-training data.
For the task-specific pre-training tasks, we select distinct hyper-parameters according to the characteristics of different downstream tasks.
Following \citet{chen2022corpusbrain}, for all task-specific pre-training tasks except entity linking and slot filling, $o$ is in [0, 1, 2, 3, 4] with a probability of [70\%, 20\%, 5\%, 3\%, 2\%], respectively.
For fact checking, we set $l$ to 3 and $n$ to 10. 
For entity linking, we set $l$ to the maximum number of anchors within $d$ to explore inter-document relations as much as possible as anchors linked to other Wikipedia pages are relatively limited.
For slot filling, we set $l$ to 3 and $k$ to 1.
For open-domain question answering, we set $l$ to 3 and $n$ to 10.
For dialogue, we set $l$ to 1 and $n$ to 10.
When it comes to rehearsing old documents, we set $k$ to 1024 and set the maximum number of iterations in the K-means cluster algorithm to 20.
In the continual pre-training phase, we set the learning rate to $1e^{-5}$ a warmup technique with a warmup ratio of 0.1.

At inference time, we use constrained beam search with 10 beams and set the maximum decoding steps to 15.
As for the entity linking sub-task, we limit the input sequence to a maximum of 384 tokens by truncating either the left, right, or both parts of the context surrounding an entity mention.

As for the generative IR baselines, i.e., DSI++ and CLEVER, we reimplement them following the empirical settings specified in the original publications since the source code has not yet been released.
To facilitate a fair comparison, we adopt the $BART_{large}$ model architecture for both models consistent with CorpusBrain++ in our implementation.

\subsection{Evaluation metrics} \label{metric introduction}
In order to assess the retrieval performance of distinct models on the KILT++ test set, as recommended in the official instructions of KILT~\cite{petroni2020kilt}, we adopt R-precision (\%) as the specific evaluation metric, i.e., the $g(\cdot)$ function in Eq.~\ref{performance}. R-precision is defined as follows,
\begin{equation}
    R\text{-}precision=\frac{r}{R},
\end{equation}
where $R$ refers to the number of Wikipedia pages in the golden provenance set, and $r$ denotes the number of relevant documents present within the top-$R$ retrieved pages.

\section{Experimental results}
In this section, we compare CorpusBrain++, baselines and variants in the dynamic retrieval scenario.
The aim is to assess the effectiveness of CorpusBrain++ in comparison to existing baselines and ablation variants.
We then evaluate CorpusBrain++ from a task-specific perspective, considering the retrieval performance in distinct tasks. 
We also evaluate the catastrophic forgetting phenomenon and forward transfer capability of distinct models.
Additionally, we analyse the effectiveness-efficiency trade-off. 
Finally, we present a case study to provide a specific illustration of the effectiveness of our proposed method.

\begin{table}[t]
\caption{Dynamic retrieval performance on individual downstream KILT datasets. We evaluate the performance of $Q_0^\sigma$,$\dots$, $Q_4^\sigma$ in terms of $VP$.} 
\label{all}
   \centering
    \setlength{\tabcolsep}{1pt}
\begin{tabular}{clcccccccccccc}
\toprule
\textbf{Session} &\textbf{Model}      &\textbf{FEV}   &\textbf{AY2} &\textbf{WnWi} &\textbf{WnCw} &\textbf{T-REx} &\textbf{zsRE} &\textbf{NQ} &\textbf{HoPo} &\textbf{TQA} &\textbf{ELI5} &\textbf{WoW} \\ \midrule
\multirow{2}{*}[-0.6cm]{0} &BM25          &32.58 	&\phantom{0}4.00 	&\phantom{0}8.77 	&\phantom{0}4.14 	&56.82 	&60.51 	&30.75 	&\textbf{42.50} 	&32.52 	&\phantom{0}6.98 	&24.10 	 \\
 &DPR &49.96 &\phantom{0}2.95 & \phantom{0}0.97 &\phantom{0}0.30 &12.00 &23.67 &45.23 &24.90 &46.56 &\textbf{15.57} &27.98             \\
 &DSI++ &\phantom{0}6.96 &56.06 &\phantom{0}9.84 &22.86 &\phantom{0}0.07 &\phantom{0}0.58 &22.32 &\phantom{0}1.58 &25.27 &\phantom{0}2.95 &12.72  \\
 &CLEVER &15.5\phantom{0} &67.93 &17.93 &28.54 &\phantom{0}1.91 &\phantom{0}6.42 &39.1\phantom{0} &\phantom{0}4.36 &38.26 &\phantom{0}4.97 &19.29  \\
 &CorpusBrain++ &\textbf{85.57}	&\textbf{83.32}	&\textbf{54.24}	&\textbf{53.95}	&\textbf{84.11}	&\textbf{97.53}	&\textbf{51.22}	&40.52	&\textbf{58.85}	&\phantom{0}9.4\phantom{0}	&\textbf{38.16}	
 \\ 
\midrule 
\multirow{2}{*}[-1.05cm]{1} &BM25 &31.17 &\phantom{0}5.79 	&\phantom{0}1.11 	&\phantom{0}3.22 	&53.69 	&58.64 	&27.61 	&41.11 	&27.55 	&\phantom{0}8.14 	&18.16 	 \\ 
&DPR  &48.46 &\phantom{0}2.89 &\phantom{0}1.11 &\phantom{0}0.77 &11.07 &24.08 &\textbf{40.40} &24.12 &46.41 &\textbf{12.79} &27.88 \\
&DSI++ &\phantom{0}0.15 &\phantom{0}0.72 &\phantom{0}0.0\phantom{0} &\phantom{0}0.46 &\phantom{0}0.2\phantom{0} &\phantom{0}0.0\phantom{0} &\phantom{0}1.35 &\phantom{0}0.07 &\phantom{0}4.67 &\phantom{0}1.16 &\phantom{0}0.26 \\
&CLEVER &\phantom{0}0.55 &\phantom{0}0.18 &\phantom{0}0.0\phantom{0} &\phantom{0}0.15 &\phantom{0}0.82 &\phantom{0}0.0\phantom{0} &\phantom{0}4.04 &\phantom{0}0.2\phantom{0} &\phantom{0}1.34 &\phantom{0}0.0\phantom{0} &\phantom{0}7.93 \\
&Direct &76.03 	&\textbf{52.26} 	&47.22 	&34.05 	&77.66 	&94.90 	&24.92 	&37.87 	&45.74 	&\phantom{0}4.65 	&31.20   \\ 
&Sequential &68.45 	&\phantom{0}8.50 	&\phantom{0}3.89 	&\phantom{0}6.60 	&70.29 	&77.90 	&19.87 	&34.91 	&39.07 	&\textbf{12.79} 	&\phantom{0}7.93 	  \\ 
&CorpusBrain++ &\textbf{77.14} &49.37 &\textbf{59.72} &\textbf{36.2}\phantom{0} &\textbf{78.07} &\textbf{96.03} &25.93 &\textbf{41.44} &\textbf{48.58} &\phantom{0}4.65 &\textbf{39.39}  \\ 
\midrule
\multirow{2}{*}[-1.05cm]{2} &BM25 &28.84 	&\phantom{0}0.48	&\phantom{0}0.30 	&\phantom{0}4.35 	&50.79 	&56.99	&28.69 	&39.10 	&27.87 	&\phantom{0}4.52 	&\textbf{10.69} 	 \\
&DPR &44.36 &\phantom{0}2.15 &\phantom{0}1.81 &\phantom{0}0.87 &\phantom{0}7.74 &20.58 &\textbf{43.18} &24.73 &44.95 &\textbf{14.57} &\phantom{0}3.05\\
&DSI++ &\phantom{0}0.37 &\phantom{0}0.0\phantom{0} &\phantom{0}0.6\phantom{0} &\phantom{0}0.22 &\phantom{0}0.0\phantom{0} &\phantom{0}0.0\phantom{0} &\phantom{0}3.34 &\phantom{0}0.06 &\phantom{0}2.73 &\phantom{0}0.5\phantom{0} &\phantom{0}0.0\phantom{0} \\
&CLEVER &\phantom{0}1.52 &\phantom{0}0.0\phantom{0} &\phantom{0}0.0\phantom{0} &\phantom{0}0.0\phantom{0} &\phantom{0}0.6\phantom{0} &\phantom{0}0.0\phantom{0} &\phantom{0}1.67 &\phantom{0}0.6\phantom{0} &\phantom{0}0.27 &\phantom{0}0.5\phantom{0} &\phantom{0}0.0\phantom{0}\\
&Direct &75.70 	&47.37 	&43.07	&35.43 	&74.01 	&96.04 	&27.30 	&38.68 	&41.80 	&\phantom{0}5.03 	&\phantom{0}9.16 	 \\
&Sequential &70.53 	&\phantom{0}4.07 	&\phantom{0}4.22 	&\phantom{0}8.26 	&55.95 	&63.32 	&13.37 	&29.70 	&32.38 	&\phantom{0}8.54 	&\phantom{0}7.63 	 \\
&CorpusBrain++ &\textbf{80.64} &\textbf{49.28} &\textbf{62.05} &\textbf{38.48} &\textbf{74.21} &\textbf{96.83} &31.2\phantom{0} &\textbf{43.65} &\textbf{48.63} &\phantom{0}7.54 &\phantom{0}7.63  \\ 
\midrule
\multirow{2}{*}[-1.05cm]{3}&BM25 &30.19 	&\phantom{0}5.67 	&\phantom{0}0.35 	&\phantom{0}3.80 	&48.10 	&57.07 	&26.52 	&37.9\phantom{0}	&31.07 	&\phantom{0}3.98 	&17.59 	 \\
&DPR &38.39 &\phantom{0}5.67 &\phantom{0}0.69 &\phantom{0}0.00 &11.21 &21.34 &\textbf{38.64} &22.89 &47.18 &\textbf{12.94} &\textbf{23.62}\\
&DSI++ &\phantom{0}0.43 &\phantom{0}0.0\phantom{0} &\phantom{0}0.0\phantom{0} &\phantom{0}0.33 &\phantom{0}0.0\phantom{0} &\phantom{0}0.0\phantom{0} &\phantom{0}1.77 &\phantom{0}0.05 &\phantom{0}1.69 &\phantom{0}1.0\phantom{0} &\phantom{0}0.0\phantom{0}  \\
&CLEVER &\phantom{0}0.99 &\phantom{0}0.0\phantom{0} &\phantom{0}0.35 &\phantom{0}0.0\phantom{0} &\phantom{0}0.18 &\phantom{0}1.44 &\phantom{0}2.53 &\phantom{0}0.1\phantom{0} &\phantom{0}0.56 &\phantom{0}0.5\phantom{0} &\phantom{0}0.0\phantom{0}\\
&Direct &64.15 	&56.92 	&44.44 	&38.51 	&74.50 	&95.20 	&28.03 	&35.71 	&45.76 	&\phantom{0}8.96 	&16.08  \\
&Sequential &57.04 	&\phantom{0}9.52 	&\phantom{0}3.82 	&16.36 	&47.02 	&62.35 	&\phantom{0}9.60 	&27.22 	&34.46 	&\phantom{0}8.46 	&19.10  \\
&CorpusBrain++ &\textbf{69.24} &\textbf{58.05} &\textbf{65.62} &\textbf{39.83} &\textbf{73.78} &\textbf{96.88} &29.8\phantom{0} &\textbf{40.72} &\textbf{50.71} &\phantom{0}6.47 &19.6\phantom{0}  \\ 
\midrule
\multirow{2}{*}[-1.05cm]{4}&BM25 &27.74 	&\phantom{0}1.90 	&\phantom{0}0.00 	&\phantom{0}2.00 	&44.73 	&46.99 	&25.93 	&38.60 	&30.24 	&\phantom{0}9.47 	&24.56	 \\
&DPR &43.65 &\phantom{0}2.48 &\phantom{0}0.27 &\phantom{0}0.00 &\phantom{0}9.67 &21.78 &\textbf{38.19} &21.68 &44.15 &\textbf{13.16} &23.39  \\
&DSI++ &\phantom{0}1.78 &\phantom{0}0.38 &\phantom{0}0.0\phantom{0} &\phantom{0}0.0\phantom{0} &\phantom{0}0.0\phantom{0} &\phantom{0}0.0\phantom{0} &\phantom{0}0.46 &\phantom{0}0.14 &\phantom{0}1.16 &\phantom{0}0.0\phantom{0} &\phantom{0}0.0\phantom{0} \\
&CLEVER &\phantom{0}1.01 &\phantom{0}0.0\phantom{0} &\phantom{0}0.0\phantom{0} &\phantom{0}0.0\phantom{0} &\phantom{0}0.17 &\phantom{0}0.86 &\phantom{0}1.85 &\phantom{0}0.09 &\phantom{0}0.12 &\phantom{0}0.0\phantom{0} &\phantom{0}0.0\phantom{0}\\
&Direct &66.57 	&\textbf{53.14} 	&47.53 	&34.93 	&\textbf{66.67} 	&95.13 &23.61 	&35.47 	&48.09	&\phantom{0}1.58 	&26.90 	\\
&Sequential &61.35 	&\phantom{0}2.29 	&\phantom{0}4.40 	&\phantom{0}5.79 	&38.69 	&53.58 	&\phantom{0}9.95 	&25.75 	&29.90 	&\phantom{0}7.37 	&\phantom{0}8.19 	 \\
&CorpusBrain++ &\textbf{71.98} &50.86 &\textbf{67.03} &\textbf{39.72} &\textbf{66.67} &\textbf{97.13} &23.84 &\textbf{40.47} &\textbf{53.42} &\phantom{0}2.11 &\textbf{32.16}  \\
\bottomrule
\vspace{-3mm}
\end{tabular}
\end{table}

\subsection{Baseline comparison}
CorpusBrain has been shown to outperform traditional IR methods such as BM25 and DPR as well as generative IR methods such as GENRE and SEAL in the non-incremental retrieval scenario~\cite{chen2022corpusbrain}. 
To streamline our work, we do not repeat the experiments and refer, instead, to \cite{chen2022corpusbrain} for more empirical details.
In this work, we solely evaluate the retrieval performance of distinct models in the incremental scenario.
In the following, we first analyze the incremental effectiveness of distinct models on eleven individual downstream datasets.
Additionally, we analyze the incremental effectiveness of distinct models on five individual downstream tasks,
each of which contains one to four datasets.
Finally, we analyze the overall incremental effectiveness of distinct models on all downstream datasets.

\subsubsection{Incremental performance of individual downstream datasets}
\autoref{all} provides an overview of the incremental retrieval performance of different models on specific KILT datasets. 
When we look at the retrieval results presented in \autoref{all}, we observe the following:
\begin{enumerate}[label=(\arabic*),leftmargin=*]
    \item DPR exhibits better performance than BM25 in the majority of downstream datasets, and the underlying reason may be that the supervised DPR method learns more semantic characteristics of downstream datasets than BM25, rendering it more adaptable in the face of newly arrived documents.
    However, the comparison results on the T-REx and zsRE datasets are opposite, which may be due to the significant difference between the input query format of these two datasets (phrases) and the query format learned by DPR (sentences).

    \item Generative retrieval baselines, i.e.,  DSI++ and CLEVER, exhibit retrieval ability to some degree in non-incremental scenarios, especially on the AY2 and NQ datasets.
    Nevertheless, this retrieval ability fails to be effectively sustained in incremental scenarios.
    This may be attributed to the exclusive focus of these models on homogeneous downstream queries during continual learning, while the dynamic retrieval scenario for KILTs is dominated by heterogeneous downstream queries.
    The failure of previous generative retrieval models highlights that it is a non-trivial challenge to continually model the relevant relationship between heterogeneous downstream queries and the corresponding documents. 

    \item In general, CorpusBrain++ consistently surpasses traditional retrieval methods and generative retrieval models in the majority of downstream datasets.

    \item CorpusBrain++ performs worse than DPR on the NQ dataset in incremental sessions, the underlying reason may be that NQ serves as one of the training datasets employed by DPR, rendering DPR a stronger retriever on the NQ dataset.

    \item CorpusBrain++ exhibits unstable retrieval performance on the AY2, ELI5 and WOW datasets.
    What these datasets have in common is that the input queries are all of a relatively long and complex form.
    For example, ELI5 is a dataset for long-form question answering, which contains complex and diverse questions that require explanatory multi-sentence answers~\cite{fan2019eli5}.
    More challengingly, no training data is provided for ELI5.
    This phenomenon may suggest that CorpusBrain++ is unstable in retrieving results in the face of long and complex input queries, and a future avenue is to boost the query length and complexity in the task-specific pre-training objectives.
\end{enumerate}

\begin{figure}[t]   

  \centering            
  \begin{subfigure}{0.45\linewidth}
    \includegraphics[width=\linewidth]{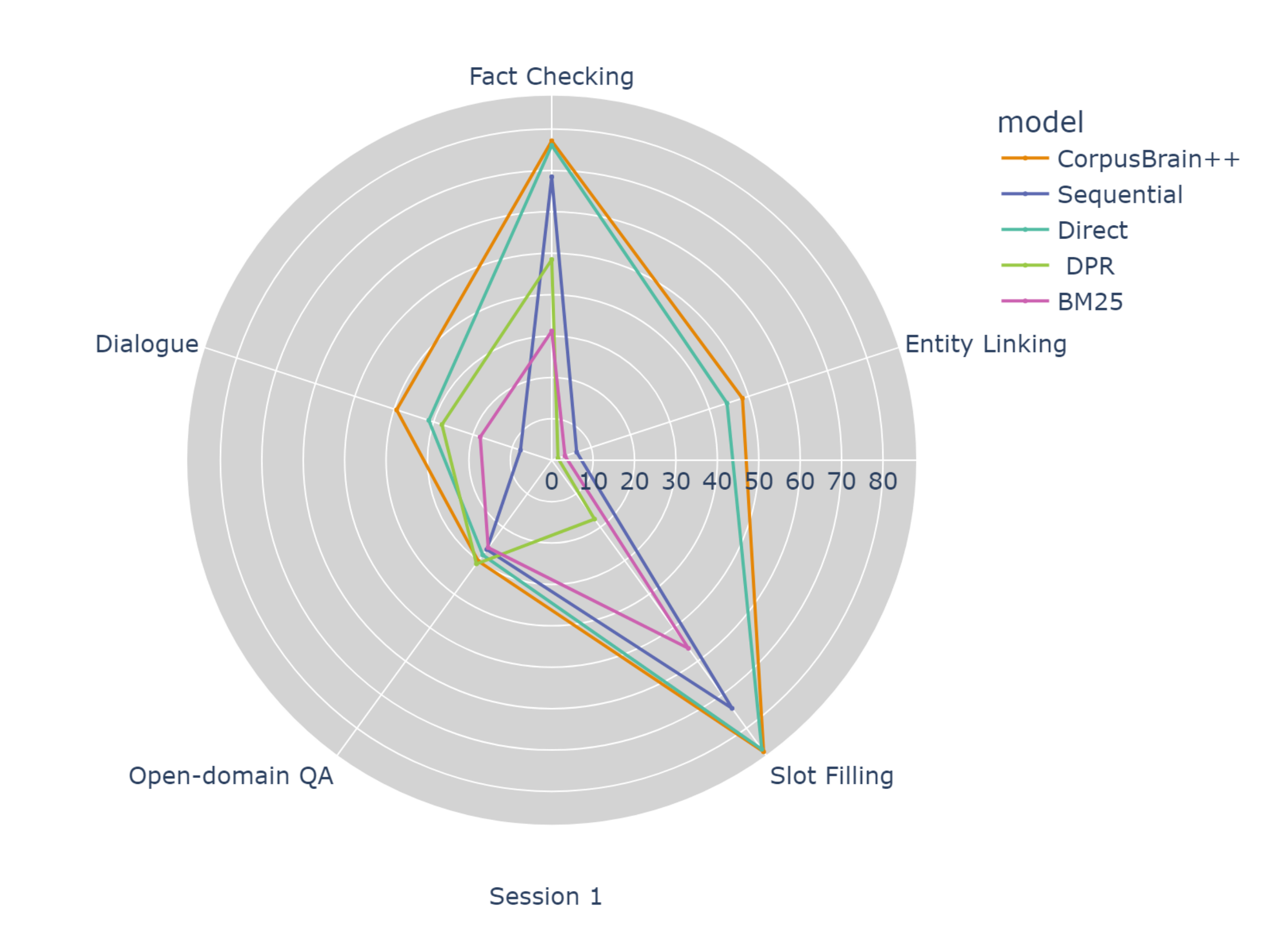}
  \end{subfigure}
\begin{subfigure}{0.45\linewidth}
    \includegraphics[width=\linewidth]{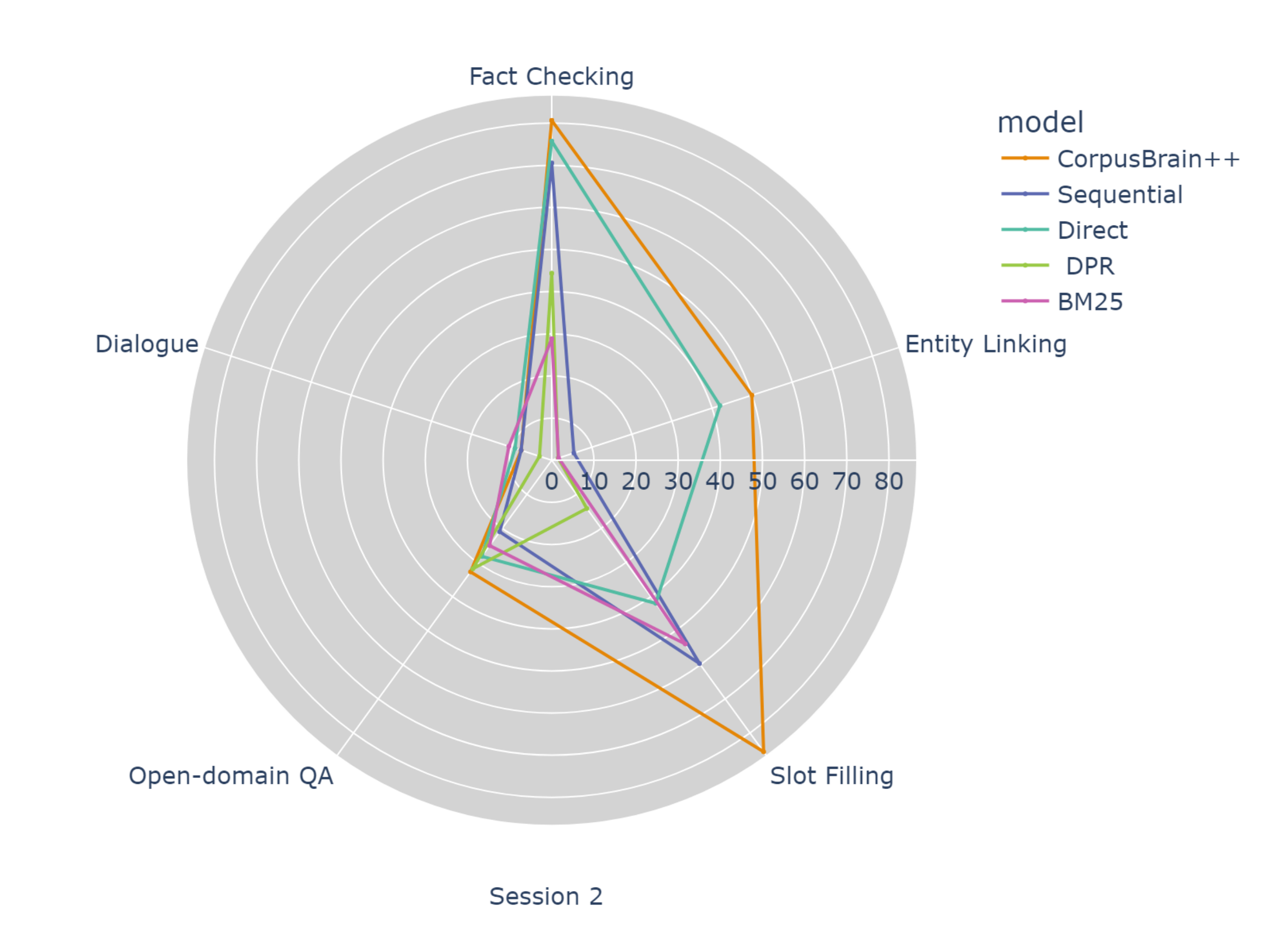}
      \end{subfigure}
  \begin{subfigure}{0.45\linewidth}
    \includegraphics[width=\linewidth]{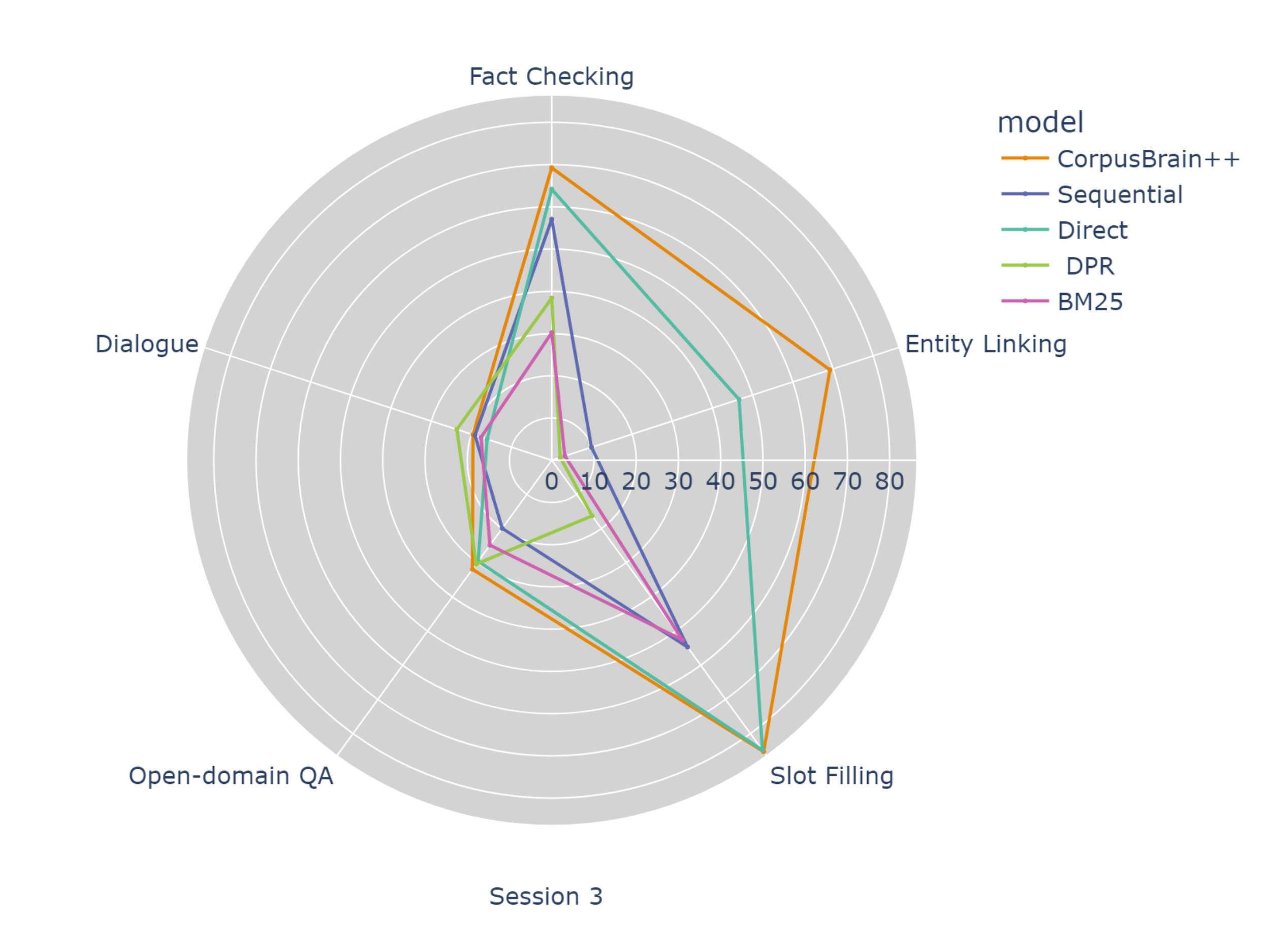}
      \end{subfigure}
  \begin{subfigure}{0.45\linewidth}
    \includegraphics[width=\linewidth]{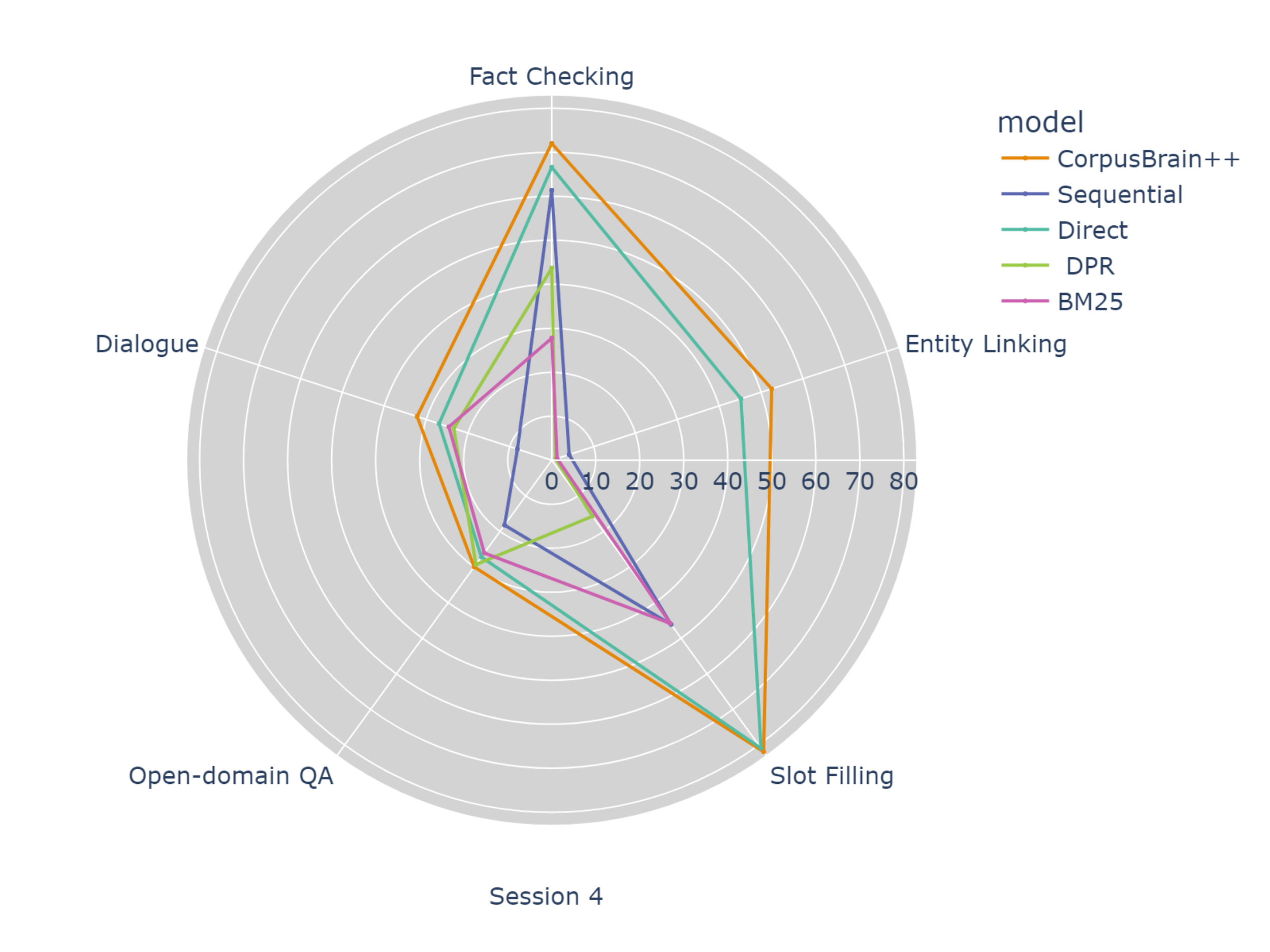}
      \end{subfigure}
  \caption{Dynamic retrieval performance of distinct models on downstream KILT tasks. We evaluate the performance of $Q_0^\tau$,$\dots$, $Q_4^\tau$ in terms of $VP$.
  }
  \label{tasks}  
\end{figure}

\subsubsection{Incremental performance of individual downstream tasks}
As illustrated in Figure \ref{tasks}, CorpusBrain++  outperforms traditional retrieval methods and naive generative variants across nearly all downstream KILT tasks during all sessions.
When we examine the enhancements achieved by CorpusBrain++ in each downstream task, we observe the following:
\begin{enumerate}[label=(\arabic*),leftmargin=*]
    \item The gain in dynamic retrieval performance is notably more significant in the fact checking and entity linking tasks. We attribute this to the fact that both of these tasks are entity-centric. Consequently, our model can effectively learn from new documents by emphasizing the entities within these documents as part of the task-specific pre-training objective.
    \item The gain in dynamic retrieval performance for the slot filling task, compared to other downstream tasks, is relatively modest. 
    This phenomenon can likely be attributed to the fact that the pre-training tasks within CorpusBrain effectively align with the downstream format of the slot filling task during the initial session, allowing the backbone model to retain a substantial retrieval capability for this specific task.
    \item The improvement in dynamic retrieval performance exhibits a relatively unstable pattern in the open-domain QA task and the dialogue task. We attribute this variability to the nature of these two tasks, which are not entity-centric. In these tasks, it is not guaranteed that the entities within the target document will appear in the input query, and hence the characteristic of downstream tasks is hard to simulate in the task-specific pre-training objective.
\end{enumerate}

\begin{table}[t]

\caption{Comprehensive retrieval performance of distinct models on $Q_i$ in terms of $VP$. As for $AP$, $BWT$ and $FWT$, $\uparrow$ indicates higher is better and $\downarrow$ indicates lower is better.
$*$ indicates statistically significant improvements over all baselines (p-value < 0.05).
}
\label{main}
   \centering
\resizebox{\textwidth}{!}{
\begin{tabular}{lcccccccc}
\toprule
\textbf{Model}      & $\mathcal{Q}_0$ & $\mathcal{Q}_1$ & $\mathcal{Q}_2$ & $\mathcal{Q}_3$ & $\mathcal{Q}_4$ & \textbf{AP}$\uparrow$ & \textbf{BWT}$\downarrow$ & \textbf{FWT}$\uparrow$ \\ 
\midrule
BM25          &27.61 &25.11 &22.97 &23.84 &22.92 &22.26 &\phantom{0}2.78 &23.71  \\
DPR &22.74 &21.82 &18.91 &20.23 &19.86 &18.49 & \phantom{0}2.77 &20.21\\
DSI++   &14.66 &\phantom{0}0.82  &\phantom{0}0.71  &\phantom{0}0.48 &\phantom{0}0.36  &\phantom{0}1.4\phantom{0} &\phantom{0}2.72 &\phantom{0}0.59 \\ 
CLEVER      &22.20 &\phantom{0}1.38 &\phantom{0}0.47 &\phantom{0}0.6\phantom{0} &\phantom{0}0.38 &\phantom{0}0.62 &\phantom{0}5.49 &\phantom{0}0.7\phantom{0}\\ 
Direct        &59.72 &47.86 &44.87 &46.21 &45.42 &48.28 &\phantom{0}0.67 &46.09 \\
Sequential   &59.72 &31.84 &27.09 &26.81 &22.48 &17.67 &19.89 &27.05   \\ 
\midrule 
$\text{CorpusBrain++}_{-Adapter(ST)}$ &59.72 &26.15 &24.66 &23.23 &22.26 &22.76 &10.55 &24.08 \\
$\text{CorpusBrain++}_{-Adapter(MT)}$ &59.72 &33.06 &33.29 &31.71 &31.78 &32.97 &\phantom{0}6.26 &32.46\\
$\text{CorpusBrain++}_{OriPT}$ &59.72 &45.33 &41.15 &40.5\phantom{0} &39.29 &41.85 &\phantom{0}4.18 &41.57\\
$\text{CorpusBrain++}_{Random}$ &59.72 &49.62 &47.05 &47.67 &47.12 &49.69 &\phantom{0}0.8\phantom{0} &47.86 \\
$\text{CorpusBrain++}_{-Rehearsal}$ &59.72 &49.26 &47.56 &47.85 &46.96 &49.55 &\phantom{0}1.05 &47.91 \\ 
\midrule 
CorpusBrain++ &\phantom{0}\textbf{59.72}$^*$ &\phantom{0}\textbf{50.59}$^*$ &\phantom{0}\textbf{49.10}$^*$ &\phantom{0}\textbf{50.06}$^*$ &\phantom{0}\textbf{49.58}$^*$ &\phantom{0}\textbf{52.01}$^*$ &\phantom{00}\textbf{0.41}$^*$ &\phantom{0}\textbf{49.83}$^*$\\
\bottomrule
\vspace{-3mm}
\end{tabular}
}
\end{table}

\subsubsection{Incremental performance of all datasets}

\autoref{main} presents an overview of the overall dynamic retrieval performance of different models.

\heading{Overall analysis}
Based on \autoref{main}, we find that:
\begin{enumerate}[label=(\arabic*),leftmargin=*]
    \item Traditional IR models, i.e., BM25 and DPR,  suffer from a modest drop of approximately 9\% and 4\% respectively in terms of VP in the first incremental session, and the observation across later sessions is consistent.
    Despite the modest decrease in retrieval capability, both traditional IR models exhibit weaker retrieval ability compared with the generative IR model CorpusBrain.

    \item Generative IR baselines, i.e., DSI++ and CLEVER, almost completely lose the retrieval capability during the incremental phase, the underlying reason may be two-fold.
    First, both baselines focus on the dynamic scenario of a single downstream task, while the dynamic retrieval scenario for KILTs comprises various downstream tasks spanning multiple semantic granularity.
    Second, Wikipedia titles, given their robust semantic structure, can effectively serve as docids in the retrieval scenario for KILTs, whereas the atomic integer docid employed in DSI++ and the product quantization code used in CLEVER may increase the difficulty of continual learning.

    \item In line with the analysis presented in Section \ref{analysis}, when naive variants of CorpusBrain, namely Direct and Sequential, are exposed to the arrival of new documents, we observe a substantial decline in retrieval performance, especially in the Sequential variant.
    This phenomenon serves as compelling evidence that off-the-shelf CorpusBrain is susceptible to catastrophic forgetting and faces challenges when adapting to the dynamic retrieval scenarios for KILTs.

    \item Taken as a whole, based on the data presented in \autoref{main}, it is evident that CorpusBrain++ consistently achieves the best retrieval performance across all metrics, including VP, AP, BWT, and FWT.
    Given the superior performance of CorpusBrain++ when compared to traditional and generative retrieval methods, it is evident that our proposed CorpusBrain++ can effectively adapt to the dynamic retrieval scenario.
\end{enumerate}

\heading{Impact of different  model architectures}
When we compare variants with different model architectures, i.e., the backbone-only architecture incorporating $\text{CorpusBrain++}_{-Adapter(ST)}$ and $\text{CorpusBrain++}_{-Adapter(MT)}$, and the backbone-adapter architecture incorporating all other CorpusBrain++ variants, we can observe that:
\begin{enumerate}[label=(\arabic*),leftmargin=*]
    \item Variants utilizing the backbone-adapter architecture consistently demonstrate superior retrieval performance across all metrics when compared to variants employing the backbone-only architecture. 
    The backbone-adapter architecture exhibits significantly lower BWT scores, indicating its enhanced ability to mitigate catastrophic forgetting.
    This phenomenon can be attributed to the inherent characteristic of the backbone-adapter architecture, which permits updates solely to fractional meta-parameters, specifically the adapter parameters, while maintaining the stability of the backbone. 
    This design choice ensures the preservation of fundamental retrieval capabilities within the backbone.

    \item The variant $\text{CorpusBrain++}_{-Adapter(ST)}$ is outperformed by $\text{CorpusBrain++}_{-Adapter(MT)}$. A potential explanation for this discrepancy lies in the training strategy. 
    During the initial session, both pre-training and fine-tuning stages adhere to the multi-task training approach, creating a divergence from the single-task training methodology employed during the incremental session. 
\end{enumerate}

\heading{Impact of different pre-training objectives}
When we look at variants with different pre-training objectives, i.e.,  $\text{CorpusBrain++}_{OriPT}$ which 
continually pre-trains the backbone-adapter architecture with the original pre-training tasks, and CorpusBrain++
which continually pre-trains the backbone-adapter architecture with our proposed task-specific pre-training objective, we can observe and analyze as follows:
\begin{enumerate}[label=(\arabic*),leftmargin=*]
    \item Despite retaining the backbone-adapter architecture to counteract catastrophic forgetting, $\text{CorpusBrain++}_{OriPT}$ exhibits retrieval performance even worse than the naive Direct variant when new documents arrive. This observation reveals that it is not feasible to directly employ general pre-training tasks to accommodate distinct downstream tasks in incremental sessions.

    \item Regarding the question of why the multi-task learning mechanism proves effective during the initial session but falters in incremental sessions, the underlying explanation may lie in the accessibility of golden query-docid pairs for downstream KILT tasks. 
    During the initial session, the availability of these golden query-docid pairs minimizes the introduction of data noise, ensuring a more stable fine-tuning stage.
    Nevertheless, during incremental sessions, we encounter a significant challenge in the form of insufficiently labeled query-docid pairs. 
    As a consequence, persisting with pre-training our model following the paradigm of multi-task learning results in the introduction of substantial levels of data noise.
    To address this challenge, we shift our approach to follow the work line of single-task learning, allowing for a more focused learning objective.  
    Thanks to the backbone-adapter architecture, assigning an adapter for each KILT task incurs minimal computational and storage overhead.
\end{enumerate}

\heading{Impact of different document rehearsal strategies}
When we focus on variants with distinct document rehearsal strategies, i.e., $\text{CorpusBrain++}_{-Rehearsal}$ without document rehearsal, $\text{CorpusBrain++}_{Random}$ with a random rehearsal strategy, and CorpusBrain++ with a document rehearsal strategy based on semantic similarity, we can observe that:
\begin{enumerate}[label=(\arabic*),leftmargin=*]
    \item Among the variants, $\text{CorpusBrain++}_{-Rehearsal}$ displays the poorest performance in terms of BWT, underscoring the efficacy of the old document rehearsal strategy in further alleviating the phenomenon of catastrophic forgetting.
    \item CorpusBrain++ demonstrates superior performance across all metrics when compared to $\text{CorpusBrain++}_{Random}$, highlighting the effectiveness of the semantic-similarity-based document rehearsal strategy.
\end{enumerate}

\subsection{Assessing catastrophic forgetting}
\begin{figure}[t]
  \centering 
  \vspace{-3mm} 
  \includegraphics[width=\linewidth]{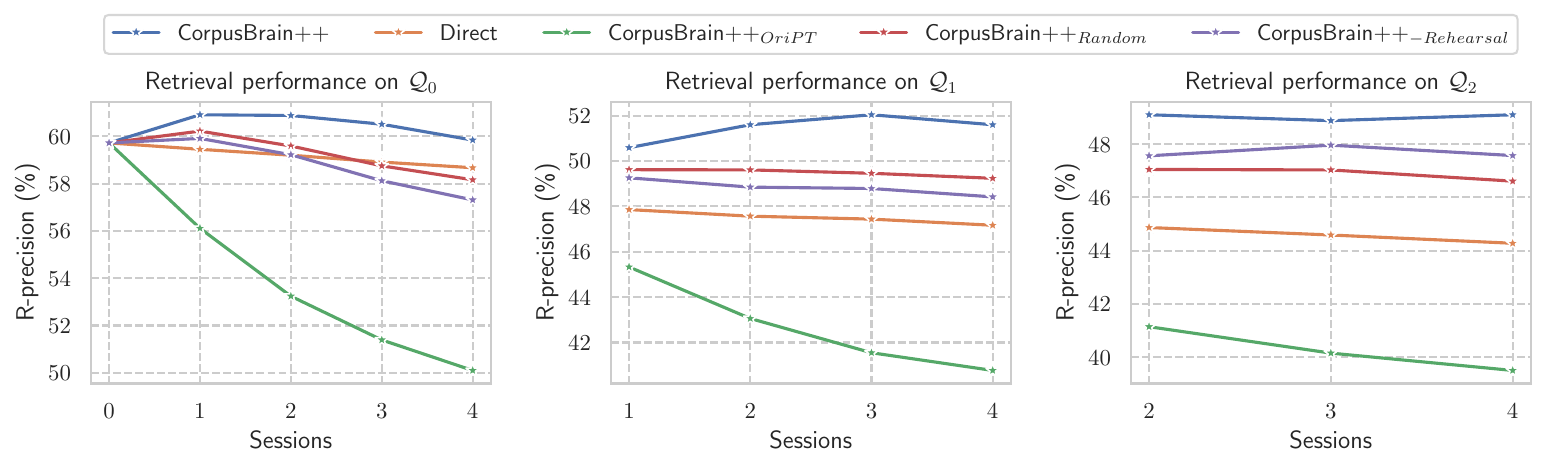}
  \caption{The catastrophic forgetting phenomenon of different models. 
  Based on the comprehensive retrieval performance of all datasets, we illustrate the retrieval performance on $Q_i$ in terms of $VP$.}
  \label{forget}
    \vspace{-3mm}
\end{figure}
In order to further assess the forgetting behavior of distinct models, we illustrate the forgetting curve of distinct models as the session grows in terms of the retrieval performance on $Q_0$, $Q_1$, and $Q_2$.
We select models with a relatively low BWT score including CorpusBrain++, Direct, $\text{CorpusBrain++}_{OriPT}$, $\text{CorpusBrain++}_{Random}$ and $\text{CorpusBrain++}_{-Rehearsal}$ for comparison.
In \autoref{forget} we observe that:
\begin{enumerate*}[label=(\roman*)]
    \item The forgetting curve for $\text{CorpusBrain++}_{OriPT}$ exhibits a notably steeper decrease compared to other models. The underlying reason might be that retrieval capabilities for downstream KILT tasks are significantly weakened in the incremental phase without the task-specific pre-training objective.
    \item Inconsistent with $Q_0$ and $Q_1$, $\text{CorpusBrain++}_{Random}$, surprisingly, even reinforces catastrophic forgetting in terms of retrieval performance on $Q_2$ compared with $\text{CorpusBrain++}_{-Rehearsal}$. This phenomenon may imply that an improper rehearsal strategy such as the random sampling strategy may even sometimes play a negative role in alleviating catastrophic forgetting.
    \item The forgetting curve for the naive variant Direct is relatively flat, which we attribute to the fact that the model parameters are constantly kept fixed in Direct.
    Influenced by the frozen parameters, we can observe that Direct exhibits relatively worse retrieval performance in the incremental phase compared with other models.
    \item CorpusBrain++ allows almost complete prevention of catastrophic forgetting, which proves the effectiveness of the task-specific pre-training objective and the semantic-similarity-based document rehearsal strategy.
\end{enumerate*}

\begin{table}[t]

\caption{The forward transferring phenomenon of different models. 
  Based on the comprehensive retrieval performance of all datasets, we illustrate the retrieval performance on $Q_i$ in terms of $VP$. $*$ indicates statistically significant improvements over all baselines (p-value < 0.05).} 
\label{forward}
   \centering
\begin{tabular}{lccccc}
\toprule
\textbf{Model}      & $\mathcal{Q}_0$ & $\mathcal{Q}_1$ & $\mathcal{Q}_2$ & $\mathcal{Q}_3$ & $\mathcal{Q}_4$ \\ \midrule
Individual &59.72 &50.59 &47.24 &47.17 &46.54 \\
CorpusBrain++ &\phantom{0}\textbf{59.72}$^*$ &\phantom{0}\textbf{50.59}$^*$ &\phantom{0}\textbf{49.10}$^*$ &\phantom{0}\textbf{50.06}$^*$ &\phantom{0}\textbf{49.58}$^*$ \\
\bottomrule
\vspace{-3mm}
\end{tabular}
\end{table}
\subsection{Assessing forward transfer}

In order to further assess the forward transfer ability of CorpusBrain++, which measures the capacity to utilize prior knowledge in adapting to new sessions, we design a new variant denoted as Individual.
In Individual, we continually pre-train the task-specific adapters individually with the tailored pre-training objective in each session, without initializing the parameters of adapters from the prior session.
Importantly, the training process for both methods during session 0 and session 1 is entirely equivalent.
As illustrated in \autoref{forward}, CorpusBrain++ consistently outperforms Individual by a substantial margin starting from session 2. This result further confirms the robust forward transfer ability of CorpusBrain++.

\subsection{Effectiveness-efficiency trade-off}

\begin{figure}[t]
  \centering 
  \vspace{-3mm} 
  \includegraphics[width=\linewidth]{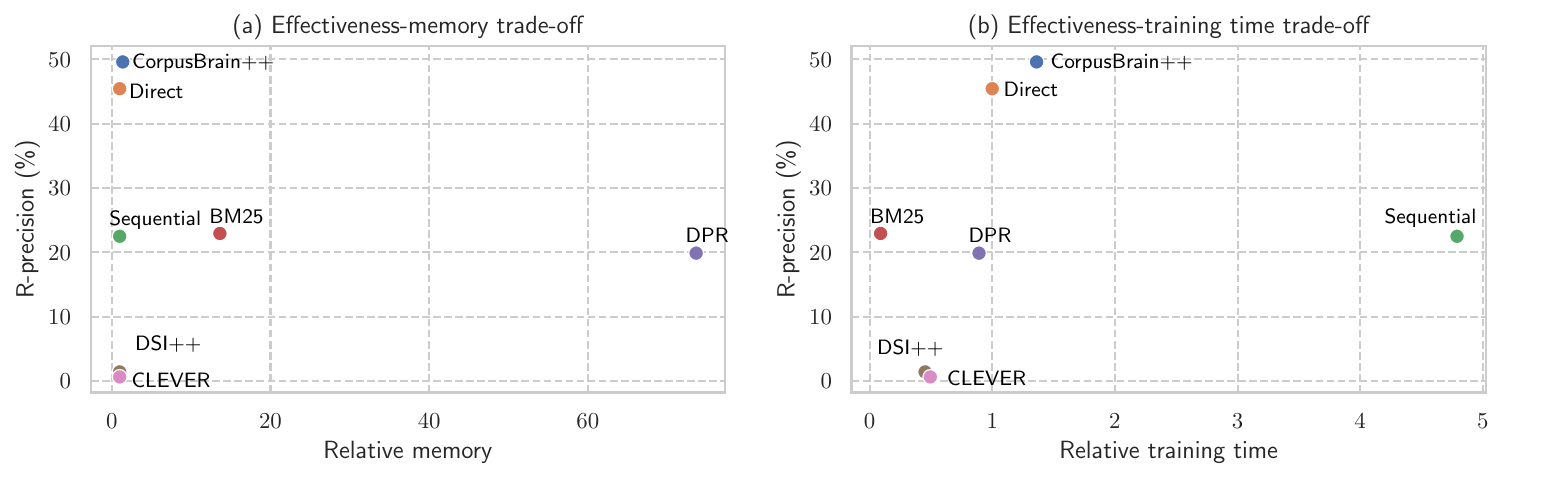}
  \caption{Comparison on effectiveness-memory trade-off and effectiveness-training time trade-off. Up and left is better. Relative memory usage and the relative training time are with respect to Direct.}
  \label{trade-off}
    \vspace{-3mm}
\end{figure}

We undertake a further comparison of the effectiveness-efficiency trade-off across various models.
Specifically, we select traditional IR methods including BM25 and DPR, as well as generative IR methods including DSI++, CLEVER, Direct, Sequential and CorpusBrain++.
As for effectiveness, we evaluate the retrieval performance on $\mathcal{Q}_T$ in terms of $VP$ after finishing training for all $T$ sessions.
For the memory overhead, we calculate the disk space usage of each model after finishing document learning of all sessions.
For the temporal overhead, we compare the total training time incurred at the conclusion of document learning.
In the case of CorpusBrain++, the task-specific adapters are continually pre-trained in parallel, hence we only count the training time of the most time-consuming adapter in the incremental phase.
As depicted in \autoref{trade-off}, both memory and training time are presented as relative ratios with respect to Direct, which enhances the clarity of the comparison.

When we look at the effectiveness-memory trade-off presented in \autoref{trade-off}(a), we observe that:
\begin{enumerate*}[label=(\roman*)]
    \item   The memory overhead of generative IR methods is significantly more modest than that of traditional IR methods, exhibiting a difference of an order of magnitude. 
    This discrepancy can be attributed to the inherent characteristics of parameterized indexes employed in generative IR methods, which demonstrate higher rates of information compression when contrasted with the external indexes utilized in traditional IR methods.
    \item   Compared to Direct, CorpusBrain++ demonstrates a significant enhancement in effectiveness while incurring only a marginal increase in storage overhead, which further demonstrates the effectiveness and efficiency of the backbone-adapter architecture.
    \item CorpusBrain++ achieves superior retrieval performance surpassing other models, while incurring only a slight increase in memory overhead compared to Direct.
\end{enumerate*}

When we look at the effectiveness-training time trade-off presented in \autoref{trade-off}(b), we observe that:
\begin{enumerate*}[label=(\roman*)]
    \item While the training process of traditional IR methods takes a shorter time than Direct, it is noteworthy that the retrieval effectiveness achieved is suboptimal. 
    \item Sequential incurs a substantial training time cost, primarily attributable to the update of all backbone parameters. Despite this investment in training time, Sequential fails to deliver satisfactory retrieval performance. 
    \item CorpusBrain++ demonstrates superior retrieval performance while incurring only a marginally higher temporal overhead compared to Direct. This implies a commendable balance between effectiveness and temporal efficiency, suggesting a high level of practicality in real-life scenarios.
\end{enumerate*}

\begin{table}[h]
  \centering
  \caption{Case study pertaining to a newly arrived document titled \emph{Nelson Mandela} in session 1, herein we present two real input queries in the test set and the corresponding retrieval documents of CorpusBrain++ in different sessions.}
  \label{case study}
  \begin{tabularx}{\textwidth}{XXX}
    \toprule
    \textbf{Wikipedia title}: Nelson Mandela \\
     \midrule
      \textbf{Text}: Nelson Mandela Nelson Rolihlahla Mandela (; ; 18 July 1918 – 5 December 2013) was a South African anti-apartheid revolutionary, political leader, and philanthropist who served as President of South Africa from 1994 to 1999. He was the country's first black head of state and the first elected in a fully representative democratic election [\dots] \\ \midrule
      \textbf{Input query 1}: (\textbf{Entity linking}) [\dots] Viljoen broke with other right-wing whites in 1994 by taking part in the country 's first all-race elections in April of that year , saying the only way to attain self-determination was by cooperating with President \texttt{[START\_ENT]} Nelson Mandela \texttt{[END\_ENT]} 's majority African National Congress [\dots] \\ 
      \textbf{Golden provenance 1}: Nelson Mandela \\
      \midrule
      \textbf{Retrieval document 1}: \\
     Session 0: Nelson Mandela 70th Birthday Tribute [SEP] Mandla Mandela \\ 
     Session 1: Nelson Mandela [SEP] Mandla Mandela \\
     Session 2: Nelson Mandela [SEP] Mandla Mandela \\
     Session 3: Nelson Mandela [SEP] Mandla Mandela \\
     Session 4: Nelson Mandela [SEP] Mandla Mandela \\
     \midrule
     \textbf{Input query 2}: (\textbf{Open-domain QA}) Who succeeded Nelson Mandela as South African president? \\ 
     \textbf{Golden provenance 2}: Nelson Mandela \\
     \midrule
     \textbf{Retrieval document 2}: \\
     Session 0: President of South Africa [SEP] Mandela: Long Walk to Freedom \\ 
     Session 1: President of South Africa [SEP] Nelson Mandela \\
     Session 2: President of South Africa [SEP] Nelson Mandela \\
     Session 3: President of South Africa [SEP] Nelson Mandela \\
     Session 4: President of South Africa [SEP] Nelson Mandela \\
    \bottomrule
  \end{tabularx}
\end{table}

\subsection{Case study}
To provide a more comprehensive understanding of CorpusBrain++, we include a case study.
In this case, a new document titled \emph{Nelson Mandela} arrives in session 1, and the specific text is depicted in \autoref{case study}.
Herein we provide two real input examples to further explain the mechanism of CorpusBrain++.
\begin{enumerate*}[label=(\roman*)]
\item In the first example, when exposed to an entity-linking query, CorpusBrain++ cannot generate the correct docid \emph{Nelson Mandela} as the top-1 candidate provenance until session 1.
Thanks to the high consistency between constructed pseudo-pairs and golden pairs, CorpusBrain++ can perform the CDL task well from session 1.
Thanks to the document rehearsal strategy used by CorpusBrain++, it is able to remember the mapping from pseudo-queries of entity linking to the docid \emph{Nelson Mandela} through all sessions.
Notably, CorpusBrain++ can retrieve documents related to \emph{Nelson Mandela} such as \emph{Mandla Mandela} (the grandson of \emph{Nelson Mandela}) in session 0, which demonstrates the robustness and generalization capabilities of the CorpusBrain++ framework.
\item In the second example, when exposed to a QA query, CorpusBrain++ can constantly generate the docid \emph{President of South Africa} related to the question since session 0.
Although the document titled \emph{President of South Africa} is not explicitly labeled as the golden provenance, we can find that its content proves beneficial to answer the given question.
After session 1, CorpusBrain++ learns the new document titled \emph{Nelson Mandela}, and retrieves \emph{Nelson Mandela} as a top-2 candidate provenance.
\end{enumerate*}
As both cases illustrate, CorpusBrain++ is capable of retrieving accurate related documents for given downstream queries and maintaining the retrieval capability without catastrophic forgetting.

\section{Related Work}

In this section, we review three lines of related work, knowledge-intensive language tasks, generative retrieval, and continual learning.

\subsection{Knowledge-intensive language tasks} 

Knowledge-intensive language tasks (KILTs) refer to a series of language tasks that require extensive and external knowledge sources such as Wikipedia. 
For instance, fact checking requires the identification of reliable pieces of evidence to establish the authenticity of a claim~\cite{thorne2018fever}, and open-domain question answering entails the need for supporting information from the knowledge base in order to provide an accurate response~\cite{kwiatkowski2019natural,yang2018hotpotqa,joshi2017triviaqa, fan2019eli5}.
To facilitate the evaluation of KILTs, a comprehensive benchmark dataset named KILT has been proposed~\cite{petroni2020kilt}, which collects eleven datasets spanning five tasks including fact checking, dialogue, slot filling, question answering, and entity linking.
Essentially, all these tasks in KILT are grounded in the same snapshot of Wikipedia. 

Practical solutions to these tasks usually involve a two-step, pipelined framework~\cite{chen2017reading,kwiatkowski2019natural,yang2018hotpotqa,joshi2017triviaqa, fan2019eli5}, including a retriever and a reader. 
Given an input query, a retriever is used to select a limited subset of relevant
information from a large knowledge source~\cite{chen2017reading,lewis2020retrieval,petroni2020kilt,robertson2009probabilistic}. 
Subsequently, a reader is applied to produce the final results by incorporating the input queries and derived support information~\cite{lewis2020bart,raffel2020exploring,lewis2020retrieval}. 
The majority of existing approaches in the retrieval component can be divided into two categories:
\begin{enumerate*}[label=(\roman*)]
    \item sparse retrieval methods that typically involve constructing an inverted index based on term-based features, and
    \item dense retrieval methods that generally construct a vectorized index based on semantic features and rely on approximate nearest neighbor search algorithms to facilitate efficient retrieval.
\end{enumerate*}
Very recently, generative retrieval methods have been proposed and employed to tackle the retrieval task for KILT~\cite{de2020autoregressive,chen2022gere, chen2022corpusbrain}.
CorpusBrain~\cite{chen2022corpusbrain} is an example of this approach; it achieves state-of-the-art retrieval performance.

The majority of prior research on KILTs, including CorpusBrain, is concentrated exclusively on static knowledge source corpus.
Nevertheless, in real-world scenarios knowledge accumulates over time, leading to an evolution of the knowledge source corpus.
Unfortunately, this pervasive scenario of a dynamic knowledge source corpus has mostly been neglected so far. 
To the best of our knowledge, our work is the first attempt to focus on the dynamic retrieval scenario for KILTs.

\subsection{Generative retrieval}

Traditional methods for IR typically involve a multi-step pipeline paradigm, i.e., the ``index-retrieve-then-rank'' paradigm \cite{dai2020context,frej2020learning,karpukhin2020dense}.
Specifically, the paradigm typically boils down to three sequential steps: 
\begin{enumerate*}[label=(\roman*)]
    \item creating an index of documents based on their content,
    \item querying the index to retrieve relevant documents, and
    \item ranking the retrieved documents based on their relevance to the query.
\end{enumerate*}
The pipeline paradigm has stood the test of time due to its adaptability and reliability across applications.
Though well-established, the pipeline paradigm encounters several challenges:
\begin{enumerate*}[label=(\roman*)]
    \item During training, heterogeneous ranking components are usually difficult to be optimized in an end-to-end way towards the global objective.
    \item During inference, an additional challenge pertains to the substantial memory resource overhead necessary for constructing and maintaining the index, which is a common dilemma not only in the inverted index of sparse retrieval models such as TF-IDF~\cite{ramos2003using} and BM25~\cite{robertson2009probabilistic}, but also in the vectorized index of dense retrieval models like DPR~\cite{karpukhin2020dense}. 
    Besides, any errors or inaccuracies introduced during a particular stage can propagate through the system and potentially impact the outcomes of subsequent stages. 
\end{enumerate*}

To address these disadvantages, generative retrieval (GR) has been proposed as an alternative paradigm.
GR refers to a new retrieval paradigm where a single consolidated model is employed to replace the commonly-used multi-stage search pipeline.
With GR, the traditional indexing stage is substituted by indexing documents into model parameters in the model training phase, and the retrieval and ranking stages are replaced by retrieving relevant documents for queries in the model inference phase~\cite{metzler2021rethinking}.
In contrast to the classic ``index-retrieve-then-rank'' paradigm, GR methods exhibit considerable advantages:
\begin{enumerate*}[label=(\roman*)]
\item GR methods allow for end-to-end optimization, hence reducing error propagation. 
\item The lack of constructing large-scale document indexes in GR reduces both time and space overhead.
\end{enumerate*}
Given the advantages, a surge of explorations of GR methods has recently emerged~\cite{de2020autoregressive, bevilacqua2022autoregressive, tay2022transformer, zhou2023enhancing}. 
GR methods mainly focus on two core issues:
\begin{enumerate*}[label=(\roman*)]
    \item how to represent documents with docids, and
    \item how to model the correlation between queries and relevant docids.
\end{enumerate*}
As for representing docids, three primary techniques are proposed, namely unstructured atomic identifiers (e.g., unique integers~\cite{tay2022transformer}), simple string identifiers (e.g., titles~\cite{de2020autoregressive,chen2022corpusbrain}), and semantically structured identifiers (e.g., clustering-based representation~\cite{tay2022transformer}).
Very recently, \citet{wang2023novo} have proposed neural optimized vocabularial docids, which are learnable by training on the retrieval tasks.
\citet{sun2023learning} have devised a document tokenization learning method to address the challenge of defining document identifiers for generative retrieval.
As for establishing the semantic mapping from documents to docids, \citet{tay2022transformer} apply memorization-based pretraining to establish a mapping between the content of documents and corresponding docids and retrieval-focused fine-tuning to facilitate the mapping of queries to relevant docids.
\citet{chen2022corpusbrain} carefully designs three pretraining tasks to generate pseudo-queries and thus resemble the relevance between downstream queries and docids.

Recently, \citet{mehta2022dsi++} have identified the challenge of catastrophic forgetting in DSI while continually indexing new documents, and have proposed DSI++, which incorporates two solutions to alleviate explicit and implicit forgetting, i.e., sharpness-aware minimization and generative memory.

In this work, we have investigated the issue of catastrophic forgetting in the context of CorpusBrain, and have devised solutions to address the problem.
Different from DSI++, we 
\begin{enumerate*}[label=(\roman*)]
    \item focus on continually pretraining rather than simply supervised fine-tuning in DSI++, and
    \item concentrate on the distinctive scenario of incremental retrieval for KILTs, where queries demonstrate a wider spectrum of diversity in contrast to the traditional retrieval scenario, incorporating varying perspectives such as task, granularity, and complexity.
\end{enumerate*}
Very recently, \citet{chen2023continual} and \citet{yoon2023continually} have also explored how to perform continual learning for generative retrieval over dynamic corpora.
Unlike their work, which mainly focuses on a single type of queries, we concentrate on incremental retrieval scenarios for KILTs spanning multiple downstream tasks.

\subsection{Continual learning}

Continual learning (CL), also commonly referred to as lifelong learning or incremental learning, is a significant and challenging research area that draws inspiration from human cognition, which tends to acquire knowledge in a sequential manner~\cite{de2021continual}.
In contrast to human beings, artificial neural networks often exhibit catastrophic forgetting when confronted with new information, leading to a loss of previously acquired knowledge~\cite{kirkpatrick2017overcoming}.
Therefore, the CL research area seeks to address this issue by exploring methods to learn from a continuous stream of data while incrementally extending existing knowledge and leveraging it for future learning.
The traditional CL scenarios~\cite{van2019three} can be grouped into three categories:
\begin{enumerate*}[label=(\roman*)]
    \item Task-incremental learning, where models are invariably equipped with task identities conveying the specific task to perform, which are distinct in different sessions.
    \item Domain-incremental learning, where the input distribution keeps changing, while the task structure remains constant in spite of unavailable task identities at test time.
    \item Class-incremental learning, where, without task identities being provided, models must be able to both solve every task seen so far and extrapolate to the tasks encountered.
\end{enumerate*}
Distinctly, our work concentrates on the continual document learning task for KILT, where the knowledge source corpus evolves over time without related KILT queries.

To avoid catastrophic forgetting of neural networks, existing methods for CL can be broadly categorized into three families: 
\begin{enumerate*}[label=(\roman*)]
    \item Replay methods~\cite{rebuffi2017icarl, shin2017continual} involve either storing previous samples in their raw format or generating pseudo-samples using a generative model. 
    While in a new session, these stored or generated samples are replayed to alleviate forgetting of previously acquired knowledge.
    \item Regularization-based methods~\cite{kirkpatrick2017overcoming, li2017learning} incorporate an additional regularization term into the loss function, which restricts the magnitude of representation change during learning on new data, hence consolidating previously acquired knowledge and alleviating catastrophic forgetting.
    \item Parameter isolation methods~\cite{mallya2018packnet, rusu2016progressive}  are typically used in  scenarios where tasks are incrementally introduced, where each task is dedicated to a unique set of model parameters independently.
\end{enumerate*}

In this work, we leverage parameter isolation methods as the primary approach to address the continual document learning task for KILTs, alongside replay methods to achieve optimal continual learning effectiveness.

\section{Conclusion and future work}
In our work, we concentrate on the dynamic retrieval scenario for knowledge-intensive language tasks (KILTs).
By defining the continual document learning (CDL) task for KILTs and introducing the new benchmark dataset KILT++, our work allows for a systematic and comprehensive assessment of the dynamic retrieval scenario for KILTs.
In particular, we present a continual generative pre-training framework for KILTs to address the CDL task.
Our framework allows for effective and efficient continual learning for KILTs, which results from a synergy between the backbone-adapter architecture and the task-specific pre-training objective tailored for each downstream KILT task.
Besides, the framework also incorporates a document rehearsal strategy based on semantic similarity to defy catastrophic forgetting of old documents.
Furthermore, a series of extensive experiments validate the effectiveness and efficiency of our framework.

\heading{Broader impact}
In practical scenarios, user queries often exhibit a broad spectrum of diversity, encompassing various perspectives such as task orientation, granularity, and complexity. 
To the best of our knowledge, we are the first to explore generative retrieval methods in the context of dynamic retrieval scenarios incorporating multiple and diverse downstream tasks.
Given that our approach accurately models real retrieval scenarios and offers effectiveness and efficiency advantages, it is well-suited for application in real-world search engines tailored to knowledge-intensive linguistic tasks.
We aim for our initial exploration to serve as a benchmark for dynamic retrieval scenarios for KILTs and to inspire the IR community to further enhance the retrieval effectiveness and efficiency in such scenarios.

\heading{Limitations and future work}
As to the limitations of our work, we currently only consider the CL paradigm of parameter isolation and experience replay.
Furthermore, the investigation of alternative CL paradigms like regularization-based methods, and the exploration of additional categories of parameter isolation methods, are both avenues that merit thorough examination.
Despite the promising results of our method, the dynamic retrieval scenario for KILTs presents several unexplored facets, particularly in the era dominated by large language models (LLMs). 
One intriguing avenue for exploration involves the design of task-specific pre-training objectives in collaboration with LLMs.

\section*{Reproducibility}
To facilitate reproducibility of the results in this paper, we have only used open datasets. The code and constructed benchmark data used to produce our results are available at  \url{https://github.com/Sherlock-coder/CorpusBrainPlusPlus}.

\begin{acks}
This work was funded by the project under Grants No. 2023YFA1011602, JCKY2022130C039 and 2021QY1701, the Youth Innovation Promotion Association CAS under Grants No. 2021100, the National Natural Science Foundation of China (NSFC) under Grants No. 62372431, and the Lenovo-CAS Joint Lab Youth Scientist Project. 
This work was also (partially) funded by 
the Hybrid Intelligence Center, a 10-year program funded by the Dutch Ministry of Education, Culture and Science through the Netherlands Organisation for Scientific Research, \url{https://hybrid-intelligence-centre.nl}, 
project LESSEN with project number NWA.1389.20.183 of the research program NWA ORC 2020/21, which is (partly) financed by the Dutch Research Council (NWO),
and
the FINDHR (Fairness and Intersectional Non-Discrimination in Human Recommendation) project that received funding from the European Union’s Horizon Europe research and innovation program under grant agreement No 101070212.

All content represents the opinion of the authors,
which is not necessarily shared or endorsed by their respective employers and/or sponsors. 
\end{acks}

\bibliographystyle{ACM-Reference-Format}
\bibliography{reference}


\begin{thebibliography}{58}


\ifx \showCODEN    \undefined \def \showCODEN     #1{\unskip}     \fi
\ifx \showDOI      \undefined \def \showDOI       #1{#1}\fi
\ifx \showISBNx    \undefined \def \showISBNx     #1{\unskip}     \fi
\ifx \showISBNxiii \undefined \def \showISBNxiii  #1{\unskip}     \fi
\ifx \showISSN     \undefined \def \showISSN      #1{\unskip}     \fi
\ifx \showLCCN     \undefined \def \showLCCN      #1{\unskip}     \fi
\ifx \shownote     \undefined \def \shownote      #1{#1}          \fi
\ifx \showarticletitle \undefined \def \showarticletitle #1{#1}   \fi
\ifx \showURL      \undefined \def \showURL       {\relax}        \fi
\providecommand\bibfield[2]{#2}
\providecommand\bibinfo[2]{#2}
\providecommand\natexlab[1]{#1}
\providecommand\showeprint[2][]{arXiv:#2}

\bibitem[Almeida et~al\mbox{.}(2007)]%
        {almeida2007evolution}
\bibfield{author}{\bibinfo{person}{Rodrigo~B. Almeida}, \bibinfo{person}{Barzan
  Mozafari}, {and} \bibinfo{person}{Junghoo Cho}.}
  \bibinfo{year}{2007}\natexlab{}.
\newblock \showarticletitle{On the Evolution of Wikipedia}. In
  \bibinfo{booktitle}{\emph{ICWSM}}.
\newblock


\bibitem[Anderson et~al\mbox{.}(2017)]%
        {anderson2017guided}
\bibfield{author}{\bibinfo{person}{Peter Anderson}, \bibinfo{person}{Basura
  Fernando}, \bibinfo{person}{Mark Johnson}, {and} \bibinfo{person}{Stephen
  Gould}.} \bibinfo{year}{2017}\natexlab{}.
\newblock \showarticletitle{Guided Open Vocabulary Image Captioning with
  Constrained Beam Search}. In \bibinfo{booktitle}{\emph{Proceedings of the
  2017 Conference on Empirical Methods in Natural Language Processing}}.
  \bibinfo{pages}{936--945}.
\newblock


\bibitem[Bevilacqua et~al\mbox{.}(2022)]%
        {bevilacqua2022autoregressive}
\bibfield{author}{\bibinfo{person}{Michele Bevilacqua},
  \bibinfo{person}{Giuseppe Ottaviano}, \bibinfo{person}{Patrick Lewis},
  \bibinfo{person}{Scott Yih}, \bibinfo{person}{Sebastian Riedel}, {and}
  \bibinfo{person}{Fabio Petroni}.} \bibinfo{year}{2022}\natexlab{}.
\newblock \showarticletitle{Autoregressive Search Engines: Generating
  Substrings as Document Identifiers}.
\newblock \bibinfo{journal}{\emph{Advances in Neural Information Processing
  Systems}}  \bibinfo{volume}{35} (\bibinfo{year}{2022}),
  \bibinfo{pages}{31668--31683}.
\newblock


\bibitem[Chen et~al\mbox{.}(2017)]%
        {chen2017reading}
\bibfield{author}{\bibinfo{person}{Danqi Chen}, \bibinfo{person}{Adam Fisch},
  \bibinfo{person}{Jason Weston}, {and} \bibinfo{person}{Antoine Bordes}.}
  \bibinfo{year}{2017}\natexlab{}.
\newblock \showarticletitle{Reading Wikipedia to Answer Open-Domain Questions}.
  In \bibinfo{booktitle}{\emph{Proceedings of the 55th Annual Meeting of the
  Association for Computational Linguistics (Volume 1: Long Papers)}}.
  \bibinfo{pages}{1870--1879}.
\newblock


\bibitem[Chen et~al\mbox{.}(2023)]%
        {chen2023continual}
\bibfield{author}{\bibinfo{person}{Jiangui Chen}, \bibinfo{person}{Ruqing
  Zhang}, \bibinfo{person}{Jiafeng Guo}, \bibinfo{person}{Maarten de Rijke},
  \bibinfo{person}{Wei Chen}, \bibinfo{person}{Yixing Fan}, {and}
  \bibinfo{person}{Xueqi Cheng}.} \bibinfo{year}{2023}\natexlab{}.
\newblock \showarticletitle{Continual Learning for Generative Retrieval over
  Dynamic Corpora}. In \bibinfo{booktitle}{\emph{Proceedings of the 32nd ACM
  International Conference on Information and Knowledge Management}}.
  \bibinfo{pages}{306--315}.
\newblock


\bibitem[Chen et~al\mbox{.}(2021)]%
        {chen2021fedmatch}
\bibfield{author}{\bibinfo{person}{Jiangui Chen}, \bibinfo{person}{Ruqing
  Zhang}, \bibinfo{person}{Jiafeng Guo}, \bibinfo{person}{Yixing Fan}, {and}
  \bibinfo{person}{Xueqi Cheng}.} \bibinfo{year}{2021}\natexlab{}.
\newblock \showarticletitle{FedMatch: Federated Learning over Heterogeneous
  Question Answering Data}. In \bibinfo{booktitle}{\emph{Proceedings of the
  30th ACM International Conference on Information \& Knowledge Management}}.
  \bibinfo{pages}{181--190}.
\newblock


\bibitem[Chen et~al\mbox{.}(2022a)]%
        {chen2022gere}
\bibfield{author}{\bibinfo{person}{Jiangui Chen}, \bibinfo{person}{Ruqing
  Zhang}, \bibinfo{person}{Jiafeng Guo}, \bibinfo{person}{Yixing Fan}, {and}
  \bibinfo{person}{Xueqi Cheng}.} \bibinfo{year}{2022}\natexlab{a}.
\newblock \showarticletitle{GERE: Generative Evidence Retrieval for Fact
  Verification}. In \bibinfo{booktitle}{\emph{Proceedings of the 45th
  International ACM SIGIR Conference on Research and Development in Information
  Retrieval}}. \bibinfo{pages}{2184--2189}.
\newblock


\bibitem[Chen et~al\mbox{.}(2022b)]%
        {chen2022corpusbrain}
\bibfield{author}{\bibinfo{person}{Jiangui Chen}, \bibinfo{person}{Ruqing
  Zhang}, \bibinfo{person}{Jiafeng Guo}, \bibinfo{person}{Yiqun Liu},
  \bibinfo{person}{Yixing Fan}, {and} \bibinfo{person}{Xueqi Cheng}.}
  \bibinfo{year}{2022}\natexlab{b}.
\newblock \showarticletitle{CorpusBrain: Pre-train a Generative Retrieval Model
  for Knowledge-Intensive Language Tasks}. In
  \bibinfo{booktitle}{\emph{Proceedings of the 31st ACM International
  Conference on Information \& Knowledge Management}}.
  \bibinfo{pages}{191--200}.
\newblock


\bibitem[Clark et~al\mbox{.}(1977)]%
        {clark1977discourse}
\bibfield{author}{\bibinfo{person}{Herbert~H Clark}, \bibinfo{person}{S
  Haviland}, {and} \bibinfo{person}{Roy~O Freedle}.}
  \bibinfo{year}{1977}\natexlab{}.
\newblock \showarticletitle{Discourse Production and Comprehension}.
\newblock In \bibinfo{booktitle}{\emph{Discourse Processes: Advances in
  Research and Theory}}. \bibinfo{publisher}{Ablex Publishing Corporation}.
\newblock


\bibitem[Clark and Haviland(1974)]%
        {clark1974psychological}
\bibfield{author}{\bibinfo{person}{Herbert~H. Clark} {and}
  \bibinfo{person}{Susan~E. Haviland}.} \bibinfo{year}{1974}\natexlab{}.
\newblock \showarticletitle{Psychological Processes as Linguistic Explanation}.
\newblock \bibinfo{journal}{\emph{Explaining Linguistic Phenomena}}
  (\bibinfo{year}{1974}), \bibinfo{pages}{91--124}.
\newblock


\bibitem[Dai and Callan(2020)]%
        {dai2020context}
\bibfield{author}{\bibinfo{person}{Zhuyun Dai} {and} \bibinfo{person}{Jamie
  Callan}.} \bibinfo{year}{2020}\natexlab{}.
\newblock \showarticletitle{Context-aware Term Weighting for First Stage
  Passage Retrieval}. In \bibinfo{booktitle}{\emph{Proceedings of the 43rd
  International ACM SIGIR Conference on Research and Development in Information
  Retrieval}}. \bibinfo{pages}{1533--1536}.
\newblock


\bibitem[De~Cao et~al\mbox{.}(2020)]%
        {de2020autoregressive}
\bibfield{author}{\bibinfo{person}{Nicola De~Cao}, \bibinfo{person}{Gautier
  Izacard}, \bibinfo{person}{Sebastian Riedel}, {and} \bibinfo{person}{Fabio
  Petroni}.} \bibinfo{year}{2020}\natexlab{}.
\newblock \showarticletitle{Autoregressive Entity Retrieval}.
\newblock \bibinfo{journal}{\emph{arXiv preprint arXiv:2010.00904}}
  (\bibinfo{year}{2020}).
\newblock


\bibitem[De~Lange et~al\mbox{.}(2021)]%
        {de2021continual}
\bibfield{author}{\bibinfo{person}{Matthias De~Lange}, \bibinfo{person}{Rahaf
  Aljundi}, \bibinfo{person}{Marc Masana}, \bibinfo{person}{Sarah Parisot},
  \bibinfo{person}{Xu Jia}, \bibinfo{person}{Ale{\v{s}} Leonardis},
  \bibinfo{person}{Gregory Slabaugh}, {and} \bibinfo{person}{Tinne
  Tuytelaars}.} \bibinfo{year}{2021}\natexlab{}.
\newblock \showarticletitle{A Continual Learning Survey: Defying Forgetting in
  Classification Tasks}.
\newblock \bibinfo{journal}{\emph{IEEE Transactions on Pattern Analysis and
  Machine Intelligence}} \bibinfo{volume}{44}, \bibinfo{number}{7}
  (\bibinfo{year}{2021}), \bibinfo{pages}{3366--3385}.
\newblock


\bibitem[Devlin et~al\mbox{.}(2019)]%
        {devlin2019bert}
\bibfield{author}{\bibinfo{person}{Jacob Devlin}, \bibinfo{person}{Ming-Wei
  Chang}, \bibinfo{person}{Kenton Lee}, {and} \bibinfo{person}{Kristina
  Toutanova}.} \bibinfo{year}{2019}\natexlab{}.
\newblock \showarticletitle{BERT: Pre-training of Deep Bidirectional
  Transformers for Language Understanding}. In
  \bibinfo{booktitle}{\emph{Proceedings of the 2019 Conference of the North
  American Chapter of the Association for Computational Linguistics: Human
  Language Technologies, Volume 1 (Long and Short Papers)}}.
  \bibinfo{pages}{4171--4186}.
\newblock


\bibitem[Fan et~al\mbox{.}(2019)]%
        {fan2019eli5}
\bibfield{author}{\bibinfo{person}{Angela Fan}, \bibinfo{person}{Yacine
  Jernite}, \bibinfo{person}{Ethan Perez}, \bibinfo{person}{David Grangier},
  \bibinfo{person}{Jason Weston}, {and} \bibinfo{person}{Michael Auli}.}
  \bibinfo{year}{2019}\natexlab{}.
\newblock \showarticletitle{ELI5: Long Form Question Answering}. In
  \bibinfo{booktitle}{\emph{Proceedings of the 57th Annual Meeting of the
  Association for Computational Linguistics}}. \bibinfo{pages}{3558--3567}.
\newblock


\bibitem[Frej et~al\mbox{.}(2020)]%
        {frej2020learning}
\bibfield{author}{\bibinfo{person}{Jibril Frej}, \bibinfo{person}{Philippe
  Mulhem}, \bibinfo{person}{Didier Schwab}, {and} \bibinfo{person}{Jean-Pierre
  Chevallet}.} \bibinfo{year}{2020}\natexlab{}.
\newblock \showarticletitle{Learning Term Discrimination}. In
  \bibinfo{booktitle}{\emph{Proceedings of the 43rd International ACM SIGIR
  Conference on Research and Development in Information Retrieval}}.
  \bibinfo{pages}{1993--1996}.
\newblock


\bibitem[Glass et~al\mbox{.}(2020)]%
        {glass2020span}
\bibfield{author}{\bibinfo{person}{Michael Glass}, \bibinfo{person}{Alfio
  Gliozzo}, \bibinfo{person}{Rishav Chakravarti}, \bibinfo{person}{Anthony
  Ferritto}, \bibinfo{person}{Lin Pan}, \bibinfo{person}{GP~Shrivatsa Bhargav},
  \bibinfo{person}{Dinesh Garg}, {and} \bibinfo{person}{Avirup Sil}.}
  \bibinfo{year}{2020}\natexlab{}.
\newblock \showarticletitle{Span Selection Pre-training for Question
  Answering}. In \bibinfo{booktitle}{\emph{Proceedings of the 58th Annual
  Meeting of the Association for Computational Linguistics}}.
  \bibinfo{pages}{2773--2782}.
\newblock


\bibitem[Graus et~al\mbox{.}(2018)]%
        {graus-birth-2018}
\bibfield{author}{\bibinfo{person}{David Graus}, \bibinfo{person}{Daan Odijk},
  {and} \bibinfo{person}{Maarten de Rijke}.} \bibinfo{year}{2018}\natexlab{}.
\newblock \showarticletitle{The Birth of Collective Memories: Analyzing
  Emerging Entities in Text Streams}.
\newblock \bibinfo{journal}{\emph{Journal of the Association for Information
  Science and Technology}} \bibinfo{volume}{69}, \bibinfo{number}{6}
  (\bibinfo{date}{June} \bibinfo{year}{2018}), \bibinfo{pages}{773--786}.
\newblock


\bibitem[Guo et~al\mbox{.}(2020)]%
        {guo2020deep}
\bibfield{author}{\bibinfo{person}{Jiafeng Guo}, \bibinfo{person}{Yixing Fan},
  \bibinfo{person}{Liang Pang}, \bibinfo{person}{Liu Yang},
  \bibinfo{person}{Qingyao Ai}, \bibinfo{person}{Hamed Zamani},
  \bibinfo{person}{Chen Wu}, \bibinfo{person}{W~Bruce Croft}, {and}
  \bibinfo{person}{Xueqi Cheng}.} \bibinfo{year}{2020}\natexlab{}.
\newblock \showarticletitle{A Deep Look into Neural Ranking Models for
  Information Retrieval}.
\newblock \bibinfo{journal}{\emph{Information Processing \& Management}}
  \bibinfo{volume}{57}, \bibinfo{number}{6} (\bibinfo{year}{2020}),
  \bibinfo{pages}{102067}.
\newblock


\bibitem[Hamerly and Elkan(2003)]%
        {hamerly2003learning}
\bibfield{author}{\bibinfo{person}{Greg Hamerly} {and} \bibinfo{person}{Charles
  Elkan}.} \bibinfo{year}{2003}\natexlab{}.
\newblock \showarticletitle{Learning the $k$ in $k$-means}.
\newblock \bibinfo{journal}{\emph{Advances in Neural Information Processing
  Systems}}  \bibinfo{volume}{16} (\bibinfo{year}{2003}).
\newblock


\bibitem[Haviland and Clark(1974)]%
        {haviland1974s}
\bibfield{author}{\bibinfo{person}{Susan~E. Haviland} {and}
  \bibinfo{person}{Herbert~H. Clark}.} \bibinfo{year}{1974}\natexlab{}.
\newblock \showarticletitle{What's New? Acquiring New Information as a Process
  in Comprehension}.
\newblock \bibinfo{journal}{\emph{Journal of Verbal Learning and Verbal
  Behavior}} \bibinfo{volume}{13}, \bibinfo{number}{5} (\bibinfo{year}{1974}),
  \bibinfo{pages}{512--521}.
\newblock


\bibitem[Houlsby et~al\mbox{.}(2019)]%
        {houlsby2019parameter}
\bibfield{author}{\bibinfo{person}{Neil Houlsby}, \bibinfo{person}{Andrei
  Giurgiu}, \bibinfo{person}{Stanislaw Jastrzebski}, \bibinfo{person}{Bruna
  Morrone}, \bibinfo{person}{Quentin De~Laroussilhe}, \bibinfo{person}{Andrea
  Gesmundo}, \bibinfo{person}{Mona Attariyan}, {and} \bibinfo{person}{Sylvain
  Gelly}.} \bibinfo{year}{2019}\natexlab{}.
\newblock \showarticletitle{Parameter-efficient Transfer Learning for NLP}. In
  \bibinfo{booktitle}{\emph{International Conference on Machine Learning}}.
  PMLR, \bibinfo{pages}{2790--2799}.
\newblock


\bibitem[Joshi et~al\mbox{.}(2017)]%
        {joshi2017triviaqa}
\bibfield{author}{\bibinfo{person}{Mandar Joshi}, \bibinfo{person}{Eunsol
  Choi}, \bibinfo{person}{Daniel~S. Weld}, {and} \bibinfo{person}{Luke
  Zettlemoyer}.} \bibinfo{year}{2017}\natexlab{}.
\newblock \showarticletitle{TriviaQA: A Large Scale Distantly Supervised
  Challenge Dataset for Reading Comprehension}. In
  \bibinfo{booktitle}{\emph{Proceedings of the 55th Annual Meeting of the
  Association for Computational Linguistics (Volume 1: Long Papers)}}.
  \bibinfo{pages}{1601--1611}.
\newblock


\bibitem[Karpukhin et~al\mbox{.}(2020)]%
        {karpukhin2020dense}
\bibfield{author}{\bibinfo{person}{Vladimir Karpukhin}, \bibinfo{person}{Barlas
  O{\u{g}}uz}, \bibinfo{person}{Sewon Min}, \bibinfo{person}{Patrick Lewis},
  \bibinfo{person}{Ledell Wu}, \bibinfo{person}{Sergey Edunov},
  \bibinfo{person}{Danqi Chen}, {and} \bibinfo{person}{Wen-tau Yih}.}
  \bibinfo{year}{2020}\natexlab{}.
\newblock \showarticletitle{Dense Passage Retrieval for Open-domain Question
  answering}.
\newblock \bibinfo{journal}{\emph{arXiv preprint arXiv:2004.04906}}
  (\bibinfo{year}{2020}).
\newblock


\bibitem[Ke et~al\mbox{.}(2020)]%
        {ke2020sentilare}
\bibfield{author}{\bibinfo{person}{Pei Ke}, \bibinfo{person}{Haozhe Ji},
  \bibinfo{person}{Siyang Liu}, \bibinfo{person}{Xiaoyan Zhu}, {and}
  \bibinfo{person}{Minlie Huang}.} \bibinfo{year}{2020}\natexlab{}.
\newblock \showarticletitle{SentiLARE: Sentiment-Aware Language Representation
  Learning with Linguistic Knowledge}. In \bibinfo{booktitle}{\emph{Proceedings
  of the 2020 Conference on Empirical Methods in Natural Language Processing
  (EMNLP)}}. \bibinfo{pages}{6975--6988}.
\newblock


\bibitem[Kingma and Ba(2014)]%
        {kingma2014adam}
\bibfield{author}{\bibinfo{person}{Diederik~P. Kingma} {and}
  \bibinfo{person}{Jimmy Ba}.} \bibinfo{year}{2014}\natexlab{}.
\newblock \showarticletitle{Adam: A Method for Stochastic Optimization}.
\newblock \bibinfo{journal}{\emph{arXiv preprint arXiv:1412.6980}}
  (\bibinfo{year}{2014}).
\newblock


\bibitem[Kirkpatrick et~al\mbox{.}(2017)]%
        {kirkpatrick2017overcoming}
\bibfield{author}{\bibinfo{person}{James Kirkpatrick}, \bibinfo{person}{Razvan
  Pascanu}, \bibinfo{person}{Neil Rabinowitz}, \bibinfo{person}{Joel Veness},
  \bibinfo{person}{Guillaume Desjardins}, \bibinfo{person}{Andrei~A Rusu},
  \bibinfo{person}{Kieran Milan}, \bibinfo{person}{John Quan},
  \bibinfo{person}{Tiago Ramalho}, \bibinfo{person}{Agnieszka
  Grabska-Barwinska}, \bibinfo{person}{Demis Hassabis},
  \bibinfo{person}{Claudia Clopath}, \bibinfo{person}{Dharshan Kumaran}, {and}
  \bibinfo{person}{Raia Hadsell}.} \bibinfo{year}{2017}\natexlab{}.
\newblock \showarticletitle{Overcoming Catastrophic Forgetting in Neural
  Networks}.
\newblock \bibinfo{journal}{\emph{Proceedings of the National Academy of
  Sciences}} \bibinfo{volume}{114}, \bibinfo{number}{13}
  (\bibinfo{year}{2017}), \bibinfo{pages}{3521--3526}.
\newblock


\bibitem[Kwiatkowski et~al\mbox{.}(2019)]%
        {kwiatkowski2019natural}
\bibfield{author}{\bibinfo{person}{Tom Kwiatkowski},
  \bibinfo{person}{Jennimaria Palomaki}, \bibinfo{person}{Olivia Redfield},
  \bibinfo{person}{Michael Collins}, \bibinfo{person}{Ankur Parikh},
  \bibinfo{person}{Chris Alberti}, \bibinfo{person}{Danielle Epstein},
  \bibinfo{person}{Illia Polosukhin}, \bibinfo{person}{Jacob Devlin},
  \bibinfo{person}{Kenton Lee}, \bibinfo{person}{Kristina Toutanova},
  \bibinfo{person}{Llion Jones}, \bibinfo{person}{Matthew Kelcey},
  \bibinfo{person}{Ming-Wei Chang}, \bibinfo{person}{Andrew~M. Dai},
  \bibinfo{person}{Jakob Uszkoreit}, \bibinfo{person}{Quoc Le}, {and}
  \bibinfo{person}{Slav Petrov}.} \bibinfo{year}{2019}\natexlab{}.
\newblock \showarticletitle{Natural Questions: A Benchmark for Question
  Answering Research}.
\newblock \bibinfo{journal}{\emph{Transactions of the Association for
  Computational Linguistics}}  \bibinfo{volume}{7} (\bibinfo{year}{2019}),
  \bibinfo{pages}{452--466}.
\newblock


\bibitem[Lewis et~al\mbox{.}(2020a)]%
        {lewis2020bart}
\bibfield{author}{\bibinfo{person}{Mike Lewis}, \bibinfo{person}{Yinhan Liu},
  \bibinfo{person}{Naman Goyal}, \bibinfo{person}{Marjan Ghazvininejad},
  \bibinfo{person}{Abdelrahman Mohamed}, \bibinfo{person}{Omer Levy},
  \bibinfo{person}{Veselin Stoyanov}, {and} \bibinfo{person}{Luke
  Zettlemoyer}.} \bibinfo{year}{2020}\natexlab{a}.
\newblock \showarticletitle{BART: Denoising Sequence-to-Sequence Pre-training
  for Natural Language Generation, Translation, and Comprehension}. In
  \bibinfo{booktitle}{\emph{Proceedings of the 58th Annual Meeting of the
  Association for Computational Linguistics}}. \bibinfo{pages}{7871--7880}.
\newblock


\bibitem[Lewis et~al\mbox{.}(2020b)]%
        {lewis2020retrieval}
\bibfield{author}{\bibinfo{person}{Patrick Lewis}, \bibinfo{person}{Ethan
  Perez}, \bibinfo{person}{Aleksandra Piktus}, \bibinfo{person}{Fabio Petroni},
  \bibinfo{person}{Vladimir Karpukhin}, \bibinfo{person}{Naman Goyal},
  \bibinfo{person}{Heinrich K{\"u}ttler}, \bibinfo{person}{Mike Lewis},
  \bibinfo{person}{Wen-tau Yih}, \bibinfo{person}{Tim Rockt{\"a}schel},
  \bibinfo{person}{Sebastian Riedel}, {and} \bibinfo{person}{Douwe Kiela}.}
  \bibinfo{year}{2020}\natexlab{b}.
\newblock \showarticletitle{Retrieval-Augmented Generation for
  Knowledge-Intensive NLP Tasks}.
\newblock \bibinfo{journal}{\emph{Advances in Neural Information Processing
  Systems}}  \bibinfo{volume}{33} (\bibinfo{year}{2020}),
  \bibinfo{pages}{9459--9474}.
\newblock


\bibitem[Li and Hoiem(2017)]%
        {li2017learning}
\bibfield{author}{\bibinfo{person}{Zhizhong Li} {and} \bibinfo{person}{Derek
  Hoiem}.} \bibinfo{year}{2017}\natexlab{}.
\newblock \showarticletitle{Learning without Forgetting}.
\newblock \bibinfo{journal}{\emph{IEEE Transactions on Pattern Analysis and
  Machine Intelligence}} \bibinfo{volume}{40}, \bibinfo{number}{12}
  (\bibinfo{year}{2017}), \bibinfo{pages}{2935--2947}.
\newblock


\bibitem[Lopez-Paz and Ranzato(2017)]%
        {lopez2017gradient}
\bibfield{author}{\bibinfo{person}{David Lopez-Paz} {and}
  \bibinfo{person}{Marc'Aurelio Ranzato}.} \bibinfo{year}{2017}\natexlab{}.
\newblock \showarticletitle{Gradient Episodic Memory for Continual Learning}.
\newblock \bibinfo{journal}{\emph{Advances in Neural Information Processing
  Systems}}  \bibinfo{volume}{30} (\bibinfo{year}{2017}).
\newblock


\bibitem[Mallya and Lazebnik(2018)]%
        {mallya2018packnet}
\bibfield{author}{\bibinfo{person}{Arun Mallya} {and} \bibinfo{person}{Svetlana
  Lazebnik}.} \bibinfo{year}{2018}\natexlab{}.
\newblock \showarticletitle{PackNet: Adding Multiple Tasks to a Single Network
  by Iterative Pruning}. In \bibinfo{booktitle}{\emph{Proceedings of the IEEE
  Conference on Computer Vision and Pattern Recognition}}.
  \bibinfo{pages}{7765--7773}.
\newblock


\bibitem[Mehta et~al\mbox{.}(2022)]%
        {mehta2022dsi++}
\bibfield{author}{\bibinfo{person}{Sanket~Vaibhav Mehta}, \bibinfo{person}{Jai
  Gupta}, \bibinfo{person}{Yi Tay}, \bibinfo{person}{Mostafa Dehghani},
  \bibinfo{person}{Vinh~Q Tran}, \bibinfo{person}{Jinfeng Rao},
  \bibinfo{person}{Marc Najork}, \bibinfo{person}{Emma Strubell}, {and}
  \bibinfo{person}{Donald Metzler}.} \bibinfo{year}{2022}\natexlab{}.
\newblock \showarticletitle{DSI++: Updating Transformer Memory with New
  Documents}.
\newblock \bibinfo{journal}{\emph{arXiv preprint arXiv:2212.09744}}
  (\bibinfo{year}{2022}).
\newblock


\bibitem[Metzler et~al\mbox{.}(2021)]%
        {metzler2021rethinking}
\bibfield{author}{\bibinfo{person}{Donald Metzler}, \bibinfo{person}{Yi Tay},
  \bibinfo{person}{Dara Bahri}, {and} \bibinfo{person}{Marc Najork}.}
  \bibinfo{year}{2021}\natexlab{}.
\newblock \showarticletitle{Rethinking Search: Making Domain Experts Out of
  Dilettantes}. In \bibinfo{booktitle}{\emph{ACM SIGIR Forum}},
  Vol.~\bibinfo{volume}{55}. ACM New York, NY, USA, \bibinfo{pages}{1--27}.
\newblock


\bibitem[Myung(2003)]%
        {myung2003tutorial}
\bibfield{author}{\bibinfo{person}{In~Jae Myung}.}
  \bibinfo{year}{2003}\natexlab{}.
\newblock \showarticletitle{Tutorial on Maximum Likelihood Estimation}.
\newblock \bibinfo{journal}{\emph{Journal of Mathematical Psychology}}
  \bibinfo{volume}{47}, \bibinfo{number}{1} (\bibinfo{year}{2003}),
  \bibinfo{pages}{90--100}.
\newblock


\bibitem[Petroni et~al\mbox{.}(2020)]%
        {petroni2020kilt}
\bibfield{author}{\bibinfo{person}{Fabio Petroni}, \bibinfo{person}{Aleksandra
  Piktus}, \bibinfo{person}{Angela Fan}, \bibinfo{person}{Patrick Lewis},
  \bibinfo{person}{Majid Yazdani}, \bibinfo{person}{Nicola De~Cao},
  \bibinfo{person}{James Thorne}, \bibinfo{person}{Yacine Jernite},
  \bibinfo{person}{Vladimir Karpukhin}, \bibinfo{person}{Jean Maillard},
  \bibinfo{person}{Vassilis Plachouras}, \bibinfo{person}{Tim Rocktäschel},
  {and} \bibinfo{person}{Sebastian Riedel}.} \bibinfo{year}{2020}\natexlab{}.
\newblock \showarticletitle{KILT: A Benchmark for Knowledge Intensive Language
  Tasks}.
\newblock \bibinfo{journal}{\emph{arXiv preprint arXiv:2009.02252}}
  (\bibinfo{year}{2020}).
\newblock


\bibitem[Raffel et~al\mbox{.}(2020)]%
        {raffel2020exploring}
\bibfield{author}{\bibinfo{person}{Colin Raffel}, \bibinfo{person}{Noam
  Shazeer}, \bibinfo{person}{Adam Roberts}, \bibinfo{person}{Katherine Lee},
  \bibinfo{person}{Sharan Narang}, \bibinfo{person}{Michael Matena},
  \bibinfo{person}{Yanqi Zhou}, \bibinfo{person}{Wei Li}, {and}
  \bibinfo{person}{Peter~J Liu}.} \bibinfo{year}{2020}\natexlab{}.
\newblock \showarticletitle{Exploring the Limits of Transfer Learning with a
  Unified Text-to-Text Transformer}.
\newblock \bibinfo{journal}{\emph{Journal of Machine Learning Research}}
  \bibinfo{volume}{21} (\bibinfo{year}{2020}), \bibinfo{pages}{1--67}.
\newblock


\bibitem[Ramos(2003)]%
        {ramos2003using}
\bibfield{author}{\bibinfo{person}{Juan Ramos}.}
  \bibinfo{year}{2003}\natexlab{}.
\newblock \showarticletitle{Using TF-IDF to Determine Word Relevance in
  Document Queries}. In \bibinfo{booktitle}{\emph{Proceedings of the first
  Instructional Conference on Machine Learning}}, Vol.~\bibinfo{volume}{242}.
  \bibinfo{pages}{29--48}.
\newblock


\bibitem[Rebuffi et~al\mbox{.}(2017)]%
        {rebuffi2017icarl}
\bibfield{author}{\bibinfo{person}{Sylvestre-Alvise Rebuffi},
  \bibinfo{person}{Alexander Kolesnikov}, \bibinfo{person}{Georg Sperl}, {and}
  \bibinfo{person}{Christoph~H. Lampert}.} \bibinfo{year}{2017}\natexlab{}.
\newblock \showarticletitle{iCaRL: Incremental Classifier and Representation
  Learning}. In \bibinfo{booktitle}{\emph{Proceedings of the IEEE Conference on
  Computer Vision and Pattern Recognition}}. \bibinfo{pages}{2001--2010}.
\newblock


\bibitem[Robertson and Zaragoza(2009)]%
        {robertson2009probabilistic}
\bibfield{author}{\bibinfo{person}{Stephen Robertson} {and}
  \bibinfo{person}{Hugo Zaragoza}.} \bibinfo{year}{2009}\natexlab{}.
\newblock \showarticletitle{The Probabilistic Relevance Framework: BM25 and
  Beyond}.
\newblock \bibinfo{journal}{\emph{Foundations and Trends in Information
  Retrieval}} \bibinfo{volume}{3}, \bibinfo{number}{4} (\bibinfo{year}{2009}),
  \bibinfo{pages}{333--389}.
\newblock


\bibitem[Rusu et~al\mbox{.}(2016)]%
        {rusu2016progressive}
\bibfield{author}{\bibinfo{person}{Andrei~A. Rusu}, \bibinfo{person}{Neil~C.
  Rabinowitz}, \bibinfo{person}{Guillaume Desjardins}, \bibinfo{person}{Hubert
  Soyer}, \bibinfo{person}{James Kirkpatrick}, \bibinfo{person}{Koray
  Kavukcuoglu}, \bibinfo{person}{Razvan Pascanu}, {and} \bibinfo{person}{Raia
  Hadsell}.} \bibinfo{year}{2016}\natexlab{}.
\newblock \showarticletitle{Progressive Neural Networks}.
\newblock \bibinfo{journal}{\emph{arXiv preprint arXiv:1606.04671}}
  (\bibinfo{year}{2016}).
\newblock


\bibitem[Shin et~al\mbox{.}(2017)]%
        {shin2017continual}
\bibfield{author}{\bibinfo{person}{Hanul Shin}, \bibinfo{person}{Jung~Kwon
  Lee}, \bibinfo{person}{Jaehong Kim}, {and} \bibinfo{person}{Jiwon Kim}.}
  \bibinfo{year}{2017}\natexlab{}.
\newblock \showarticletitle{Continual Learning with Deep Generative Replay}.
\newblock \bibinfo{journal}{\emph{Advances in Neural Information Processing
  Systems}}  \bibinfo{volume}{30} (\bibinfo{year}{2017}).
\newblock


\bibitem[Singhal(2001)]%
        {singhal2001modern}
\bibfield{author}{\bibinfo{person}{Amit Singhal}.}
  \bibinfo{year}{2001}\natexlab{}.
\newblock \showarticletitle{Modern Information Retrieval: A Brief Overview}.
\newblock \bibinfo{journal}{\emph{IEEE Data Eng. Bull.}} \bibinfo{volume}{24},
  \bibinfo{number}{4} (\bibinfo{year}{2001}), \bibinfo{pages}{35--43}.
\newblock


\bibitem[Stickland and Murray(2019)]%
        {stickland2019bert}
\bibfield{author}{\bibinfo{person}{Asa~Cooper Stickland} {and}
  \bibinfo{person}{Iain Murray}.} \bibinfo{year}{2019}\natexlab{}.
\newblock \showarticletitle{BERT and Pals: Projected Attention Layers for
  Efficient Adaptation in Multi-task Learning}. In
  \bibinfo{booktitle}{\emph{International Conference on Machine Learning}}.
  PMLR, \bibinfo{pages}{5986--5995}.
\newblock


\bibitem[Su et~al\mbox{.}(2020)]%
        {su2020continual}
\bibfield{author}{\bibinfo{person}{Lixin Su}, \bibinfo{person}{Jiafeng Guo},
  \bibinfo{person}{Ruqing Zhang}, \bibinfo{person}{Yixing Fan},
  \bibinfo{person}{Yanyan Lan}, {and} \bibinfo{person}{Xueqi Cheng}.}
  \bibinfo{year}{2020}\natexlab{}.
\newblock \showarticletitle{Continual Domain Adaptation for Machine Reading
  Comprehension}. In \bibinfo{booktitle}{\emph{Proceedings of the 29th ACM
  international conference on information \& knowledge management}}.
  \bibinfo{pages}{1395--1404}.
\newblock


\bibitem[Sun et~al\mbox{.}(2023)]%
        {sun2023learning}
\bibfield{author}{\bibinfo{person}{Weiwei Sun}, \bibinfo{person}{Lingyong Yan},
  \bibinfo{person}{Zheng Chen}, \bibinfo{person}{Shuaiqiang Wang},
  \bibinfo{person}{Haichao Zhu}, \bibinfo{person}{Pengjie Ren},
  \bibinfo{person}{Zhumin Chen}, \bibinfo{person}{Dawei Yin},
  \bibinfo{person}{Maarten de Rijke}, {and} \bibinfo{person}{Zhaochun Ren}.}
  \bibinfo{year}{2023}\natexlab{}.
\newblock \showarticletitle{Learning to Tokenize for Generative Retrieval}.
\newblock \bibinfo{journal}{\emph{arXiv preprint arXiv:2304.04171}}
  (\bibinfo{year}{2023}).
\newblock


\bibitem[Tay et~al\mbox{.}(2022)]%
        {tay2022transformer}
\bibfield{author}{\bibinfo{person}{Yi Tay}, \bibinfo{person}{Vinh Tran},
  \bibinfo{person}{Mostafa Dehghani}, \bibinfo{person}{Jianmo Ni},
  \bibinfo{person}{Dara Bahri}, \bibinfo{person}{Harsh Mehta},
  \bibinfo{person}{Zhen Qin}, \bibinfo{person}{Kai Hui}, \bibinfo{person}{Zhe
  Zhao}, \bibinfo{person}{Jai Gupta}, {et~al\mbox{.}}}
  \bibinfo{year}{2022}\natexlab{}.
\newblock \showarticletitle{Transformer Memory as a Differentiable Search
  Index}.
\newblock \bibinfo{journal}{\emph{Advances in Neural Information Processing
  Systems}}  \bibinfo{volume}{35} (\bibinfo{year}{2022}),
  \bibinfo{pages}{21831--21843}.
\newblock


\bibitem[Thorne et~al\mbox{.}(2018)]%
        {thorne2018fever}
\bibfield{author}{\bibinfo{person}{James Thorne}, \bibinfo{person}{Andreas
  Vlachos}, \bibinfo{person}{Christos Christodoulopoulos}, {and}
  \bibinfo{person}{Arpit Mittal}.} \bibinfo{year}{2018}\natexlab{}.
\newblock \showarticletitle{FEVER: a Large-scale Dataset for Fact Extraction
  and VERification}. In \bibinfo{booktitle}{\emph{Proceedings of the 2018
  Conference of the North American Chapter of the Association for Computational
  Linguistics: Human Language Technologies, Volume 1 (Long Papers)}}.
  \bibinfo{pages}{809--819}.
\newblock


\bibitem[Van~de Ven and Tolias(2019)]%
        {van2019three}
\bibfield{author}{\bibinfo{person}{Gido~M. Van~de Ven} {and}
  \bibinfo{person}{Andreas~S. Tolias}.} \bibinfo{year}{2019}\natexlab{}.
\newblock \showarticletitle{Three Scenarios for Continual Learning}.
\newblock \bibinfo{journal}{\emph{arXiv preprint arXiv:1904.07734}}
  (\bibinfo{year}{2019}).
\newblock


\bibitem[Vaswani et~al\mbox{.}(2017)]%
        {vaswani2017attention}
\bibfield{author}{\bibinfo{person}{Ashish Vaswani}, \bibinfo{person}{Noam
  Shazeer}, \bibinfo{person}{Niki Parmar}, \bibinfo{person}{Jakob Uszkoreit},
  \bibinfo{person}{Llion Jones}, \bibinfo{person}{Aidan~N Gomez},
  \bibinfo{person}{{\L}ukasz Kaiser}, {and} \bibinfo{person}{Illia
  Polosukhin}.} \bibinfo{year}{2017}\natexlab{}.
\newblock \showarticletitle{Attention is All You Need}.
\newblock \bibinfo{journal}{\emph{Advances in Neural Information Processing
  Systems}}  \bibinfo{volume}{30} (\bibinfo{year}{2017}).
\newblock


\bibitem[Wang et~al\mbox{.}(2023)]%
        {wang2023novo}
\bibfield{author}{\bibinfo{person}{Zihan Wang}, \bibinfo{person}{Yujia Zhou},
  \bibinfo{person}{Yiteng Tu}, {and} \bibinfo{person}{Zhicheng Dou}.}
  \bibinfo{year}{2023}\natexlab{}.
\newblock \showarticletitle{NOVO: Learnable and Interpretable Document
  Identifiers for Model-Based IR}. In \bibinfo{booktitle}{\emph{Proceedings of
  the 32nd ACM International Conference on Information and Knowledge
  Management}}. \bibinfo{pages}{2656--2665}.
\newblock


\bibitem[Wu et~al\mbox{.}(2020)]%
        {wu2020scalable}
\bibfield{author}{\bibinfo{person}{Ledell Wu}, \bibinfo{person}{Fabio Petroni},
  \bibinfo{person}{Martin Josifoski}, \bibinfo{person}{Sebastian Riedel}, {and}
  \bibinfo{person}{Luke Zettlemoyer}.} \bibinfo{year}{2020}\natexlab{}.
\newblock \showarticletitle{Scalable Zero-shot Entity Linking with Dense Entity
  Retrieval}. In \bibinfo{booktitle}{\emph{Proceedings of the 2020 Conference
  on Empirical Methods in Natural Language Processing (EMNLP)}}.
  \bibinfo{pages}{6397--6407}.
\newblock


\bibitem[Yang et~al\mbox{.}(2018)]%
        {yang2018hotpotqa}
\bibfield{author}{\bibinfo{person}{Zhilin Yang}, \bibinfo{person}{Peng Qi},
  \bibinfo{person}{Saizheng Zhang}, \bibinfo{person}{Yoshua Bengio},
  \bibinfo{person}{William Cohen}, \bibinfo{person}{Ruslan Salakhutdinov},
  {and} \bibinfo{person}{Christopher~D Manning}.}
  \bibinfo{year}{2018}\natexlab{}.
\newblock \showarticletitle{HotpotQA: A Dataset for Diverse, Explainable
  Multi-hop Question Answering}. In \bibinfo{booktitle}{\emph{Proceedings of
  the 2018 Conference on Empirical Methods in Natural Language Processing}}.
  \bibinfo{pages}{2369--2380}.
\newblock


\bibitem[Yoon et~al\mbox{.}(2023)]%
        {yoon2023continually}
\bibfield{author}{\bibinfo{person}{Soyoung Yoon}, \bibinfo{person}{Chaeeun
  Kim}, \bibinfo{person}{Hyunji Lee}, \bibinfo{person}{Joel Jang}, {and}
  \bibinfo{person}{Minjoon Seo}.} \bibinfo{year}{2023}\natexlab{}.
\newblock \showarticletitle{Continually Updating Generative Retrieval on
  Dynamic Corpora}.
\newblock \bibinfo{journal}{\emph{arXiv preprint arXiv:2305.18952}}
  (\bibinfo{year}{2023}).
\newblock


\bibitem[Zhang et~al\mbox{.}(2020)]%
        {zhang2020pegasus}
\bibfield{author}{\bibinfo{person}{Jingqing Zhang}, \bibinfo{person}{Yao Zhao},
  \bibinfo{person}{Mohammad Saleh}, {and} \bibinfo{person}{Peter Liu}.}
  \bibinfo{year}{2020}\natexlab{}.
\newblock \showarticletitle{Pegasus: Pre-training with Extracted Gap-sentences
  for Abstractive Summarization}. In \bibinfo{booktitle}{\emph{International
  Conference on Machine Learning}}. PMLR, \bibinfo{pages}{11328--11339}.
\newblock


\bibitem[Zhao et~al\mbox{.}(2022)]%
        {zhao2022dense}
\bibfield{author}{\bibinfo{person}{Wayne~Xin Zhao}, \bibinfo{person}{Jing Liu},
  \bibinfo{person}{Ruiyang Ren}, {and} \bibinfo{person}{Ji-Rong Wen}.}
  \bibinfo{year}{2022}\natexlab{}.
\newblock \showarticletitle{Dense Text Retrieval based on Pretrained Language
  Models: A Survey}.
\newblock \bibinfo{journal}{\emph{arXiv preprint arXiv:2211.14876}}
  (\bibinfo{year}{2022}).
\newblock


\bibitem[Zhou et~al\mbox{.}(2023)]%
        {zhou2023enhancing}
\bibfield{author}{\bibinfo{person}{Yujia Zhou}, \bibinfo{person}{Zhicheng Dou},
  {and} \bibinfo{person}{Ji-Rong Wen}.} \bibinfo{year}{2023}\natexlab{}.
\newblock \showarticletitle{Enhancing Generative Retrieval with Reinforcement
  Learning from Relevance Feedback}. In \bibinfo{booktitle}{\emph{The 2023
  Conference on Empirical Methods in Natural Language Processing}}.
\newblock


\end{thebibliography}

\end{document}